\documentclass{amsart}
\usepackage{amssymb}
\usepackage{graphicx}
\usepackage{amscd}
\usepackage{amsmath}

\theoremstyle{plain}

\numberwithin{equation}{section}

\input{tcilatex}

\begin{document}
\title[Quantum Noise, Bits and Jumps.]{Quantum Noise, Bits and Jumps:\\
Uncertainties, Decoherence, Measurements and Filtering.}
\author{V P Belavkin}
\address{Mathematics Department, University of Nottingham, NG7 2RD, UK}
\email{vpb@maths.nott.ac.uk}
\urladdr{http://www.maths.nott.ac.uk/personal/vpb/}
\thanks{Published in: Progress in Quantum Electronics 25 (2001) No 1, 1 - 53}
\date{December 20, 2000.}
\dedicatory{In celebration of the 100th anniversary of the discovery of
quanta}

\begin{abstract}
It is shown that before the rise of quantum mechanics 75 years ago, the
quantum theory had appeared first in the form of the statistics of quantum
thermal noise and quantum spontaneous jumps which have never been explained
by quantum mechanics. This led to numerous quantum paradoxes, some of them
due to the great inventors of quantum theory such as Einstein and Schr\"{o}%
dinger. They are reconsidered in this paper. The development of quantum
measurement theory, initiated by von Neumann, indicated a possibility for
resolution of this interpretational crisis by divorcing the algebra of the
dynamical generators from the algebra of the actual observables. It is shown
that within this approach quantum causality can be rehabilitated in the form
of a superselection rule for compatibility of past observables with the
potential future. This rule, together with the self-compatibility of
measurements insuring the consistency of histories, is called the
nondemolition principle. The application of this causality condition in the
form of the dynamical commutation relations leads to the derivation of the
generalized von Neumann reductions, usharp, instantaneous, spontaneous, and
even continuous in time. This gives a quantum probabilistic solution, in the
form of the dynamical filtering equations, of the notorious measurement
problem which was tackled unsuccessfully by many famous physicists starting
with Schr\"{o}dinger and Bohr. The simplest Markovian quantum stochastic
model for the continuous in time measurements involves a boundary-value
problem in second quantization for input ''offer'' waves in one extra
dimension, and a reduction of the algebra of ''actual'' observables to an
Abelian subalgebra for the output waves.
\end{abstract}

\maketitle
\tableofcontents

\section{Introduction: The Common Thread of Mathematics and Physics}

\medskip

\begin{quote}
\textit{In science one tries to tell people, in such a way as to be
understood by everyone, something that no one even knew before. But in
poetry, it's quite opposite} -- Paul Dirac.
\end{quote}

This review unravels the most obscure sides of Quantum Theory related to
quantum noise, decoherence and measurement, explaining them using the
example of quantum bits. These are the most elementary quantum systems such
as spins 1/2, which are now utilized in Quantum Information Theory and for
Quantum Computations. The vast majority of papers on these new applications
which have recently appeared in theoretical physics (see for example the
Quantum Information Section in any issue of Phys Rev A) are mainly concerned
with deterministic and mathematically well-defined quantum attributes such
as unitary evolutions and entanglements in Hilbert space which are
traditional in theoretical physics. They leave the probabilistic
`ill-defined' quantum causality, decoherence and measurement problems to
quantum philosophers for vague speculations ignoring recent mathematical
developments in solving these quantum probability questions. Surely the
words ``professional theoretical physicists ought to be able to do better''
(cited from J Bell, \cite{Bell87}, p. 173) should be also addressed to those
mathematicians and computer scientists who invent the `quantum' algorithms
simply as the entangled unitary transformations if they really want to
contribute to the prediction of the potential capabilities for Quantum
Information Technologies in the new Quantum Century.

In order to utilize the results of such purely dynamical algorithms of
quantum computations one should state rigorously and solve the dynamical
measurement, quantum prediction and feedback control problems for quantum
bits performing these computations, as it was done over 20 years ago for
linear quantum dynamical systems (oscillators) in \cite{Be80, Be99}. The
true Heisenberg principle which is explained in the last chapter of this
paper does not leave a possibility for nondemolition observation of quantum
computations without quantum noise: the\ measurement errors and dynamical
perturbations satisfy the uncertainty relation, and the best what can be
done for dynamical error correction is optimal filtering of the quantum
noise. Without the mathematical formulation of interpretational problems of
quantum theory, without use of quantum laws of large numbers in probability
and information theory, not only it is impossible to prove some speculative
conjectures about the enormous capabilities of quantum algorithms, but even
the simplest traditional quantum attributes fail to differentiate this
theory from classical theories.

Indeed, the unitary evolution and entanglement without quantum probabilistic
interpretation are also the attributes of any classical linear Hamiltonian
theory, and it is not clear why it is not possible or more practical to
realize the quantum unitary algorithms or quantum logic in the classical
form of waves and non-Boolean wave-logic in a cup of water instead of the
waves of matter. One could utilize the oscillations of the complex
amplitudes of Fourier components on the surface of the water, the
interference of the water-waves and even the entanglements of the separated
modes in the cup as the unitary evolution, interference and entanglement in
any Hilbert subspace of the wave components, and would avoid the
probabilistic uncertainties, decoherence and the measurement problem which
could destroy the advantages of such `quantum' computer.

Quantum theory is a mathematical theory which studies the most fundamental
features of reality in a unified form of waves and matter, it raises and
solves the most fundamental riddles of Nature by developing and utilizing
mathematical concepts and methods of all branches of modern mathematics,
including probability and statistics. Indeed, as we shall see, it began with
the discovery of new laws for `quantum' numbers, the natural objects \ which
are the foundation of pure mathematics. (`God made the integers; the rest is
man's work' -- Kronecker). Next it invented new applied mathematical methods
for solving quantum mechanical matrix and partial differential equations.
Next it married probability with algebra to obtain unified treatment of
waves and particles in nature, giving birth to quantum probability and
creating new branches of mathematics such as quantum logics, quantum
topologies, quantum geometries, quantum groups. It inspired the recent
creation of quantum analysis and quantum calculus, as well as quantum
statistics and quantum stochastics.

Specialists in different narrow branches of mathematics and physics rarely
understand quantum theory as a common thread which runs through everything.
The creators of quantum mechanics, the theory invented for interpretation of
the dynamical laws of fundamental particles, were unable to find a
consistent interpretation of it since they were physicists with a classical
mathematical education. After inventing quantum mechanics they spent much of
their lives trying to tackle the Problem of Quantum Measurement, the
greatest problem of quantum theory, not just of quantum mechanics, or even
of unified quantum field theory, which would be the same `thing in itself'
as quantum mechanics of closed systems without such interpretation. As we
shall see, the solution to this problem can be found in the framework of
Quantum Probability and Stochastics as a part of a unified mathematics
rather than physics. Most modern theoretical physicists have a broad
mathematical education which tends to ignore two crucial aspects of this
solution -- information theory and statistical conditioning.

In order to appreciate the quantum drama which has been developing through
the whole century, and to estimate possible consequences for it of the
solution of quantum measurement problem in the new quantum technological
age, it seems useful to give a brief account of the discovery of quantum
theory and its probabilistic interpretation at the beginning of the 20th
century. This is done in the Chapter 1 which starts with a short recall of
how the quantum nature of thermal radiation was discovered by Max Planck
giving the birth of quantum theory exactly 100 years ago \cite{Plnkqp}.
Readers who are not interested in the subject of this review from the
historical perspective are advised to start with the Chapter 2 dealing with
the famous problems and paradoxes of the orthodox quantum theory from the
modern quantum probabilistic point of view. The specialists who are familiar
with this point of view and who want only to read a systematic review on the
stochastic decoherence theory, consistent trajectories, continual
measurements, quantum jumps and filtering, or would like to find the origin
of these modern quantum theories, can go directly to the hard core of the
review in Chapter 3. Because of the celebratory nature of this article, we
refrain from pursuing the detailed implications for quantum electronics,
although there are of course many, as well as many in other areas of modern
experimental physics, see for example the recent review papers \cite{Ved98,
Bra98, Vos99} which contain many references to earlier relevant papers.

Thus, the aim of this paper is to give a comprehensive review of recent
development in modern quantum theory from the historical perspective of the
discovery the deterministic quantum evolutions by Heisenberg and Schr\"{o}%
dinger to the stochastic evolutions of quantum jumps and quantum diffusion
in quantum noise. This is the direction in which quantum theory would have
developed by the founders if the mathematics of quantum stochastics had been
discovered by that time. We shall give a brief account of this new
mathematics (which plays the same role for quantum stochastics as did the
classical differential calculus for Newtonian dynamics) and concentrate on
its application to the dynamical solution of quantum measurement problems 
\cite{Wig63, Dav76, Be78, BLP82, Hol82, Be83, Bar83}, rather than give the
full account of all related theoretical papers. Among these we would like to
mention the papers on quantum decoherence \cite{Zeh70, UnZu89}, dynamical
state reduction program \cite{Per76, Gis83}, consistent histories and
evolutions \cite{GeHa90, Haa95}, spontaneous localization and events \cite%
{GRW86, BlJa95}, restricted and unsharp path integrals \cite{Men93, AKS97}
and their numerous applications to quantum countings, jumps and trajectories
in quantum optics and atomic physics \cite{MiWa84, WCM85, Car86, ZMW87,
Bar87, HMW89, Ued90, MiGa92}. Most of these papers develop a nonstochastic
phenomenological approach which is based on a non-Hamiltonian
``instrumental'' linear master equation giving the statistics of quantum
measurements, but is not well adapted for the description of individual and
conditional behavior under the continuous measurements. Pearl and Gisen took
an opposite deterministic nonlinear approach for the individual evolutions
without considering the statistics of measurements in their earlier papers 
\cite{Per76, Gis83}. In this review we shall concentrate on a more
constructive modern stochastic approach \cite{Be80, Gis84, Be85, BaLu85,
Be88, Dio88, Be89a, Gis89, Be90b, GPR90, Be90c} which gives the output
statistics recurrently as a result of the solution of a stochastic
differential equation. This allows the direct application of quantum
conditioning and filtering methods to tackle the dynamical problem of
quantum individual dynamics under continual (trajectorial) measurement. Here
we refer mainly to the pioneering and original papers in which the relevant
quantum structures as mathematical notions and methods were first invented
in the mathematical physics literature. There are also many excellent
publications in the theoretical and applied physics literature which
appeared during the 90's such as the state diffusion theory \cite{GiPe92,
GiPe93}, and especially in quantum optics \cite{Car93, WiMi93, GoGr93,
WiMi94, GoGr94, Car94} where the nonlinear quantum stochastic equations for
continual measurements have been used. However the transition from nonlinear
to linear stochastic equations and the quantum stochastic unitary models for
the underlying Hamiltonian microscopic evolutions remain unexplained in
these papers. An exception occurred in \cite{GoGr94, GGH95}, where our
quantum stochastic theory was well understood both at a macroscopic and
microscopic level. Most of these papers treat very particular
phenomenological models which are based only on the counting, or sometimes
diffusive (homodyne and heterodyne) models of quantum noise and output
process, and reinvent many notions such as quantum conditioning with respect
to the individual trajectories, without references to the general quantum
stochastic measurement and filtering (conditioning) theory which have been
developed for these purposes in the 80's. This explains why a systematic
review of this kind is needed, and, hopefully, will be appreciated.

\section{The Quantum Century Begins}

\medskip

\begin{quote}
\textit{The whole is more than the sum of its parts} -- Aristotle.
\end{quote}

This is the famous superadditivity law from Aristotle's \textit{Metaphysics}
which studies `the most general or abstract features of reality and the
principles that have universal validity'. Certainly in this broad definition
quantum physics is the most essential part of metaphysics.

Quantum theory is one of the greatest intellectual achievements of the past
century. Since the discovery of quanta by Max Planck exactly 100 years ago
on the basis of spectral analysis of quantum thermal noise, and the wave
nature of matter, it has produced numerous paradoxes and confusions even in
the greatest scientific minds such as those of Einstein, de Broglie, Schr%
\"{o}dinger, Bell, and it still confuses many contemporary philosophers and
scientists who fail to accept the Aristotle's superadditivity law of Nature.
Rapid development of the beautiful and sophisticated mathematics for quantum
mechanics and the development of its interpretation by Bohr, Born,
Heisenberg, Dirac, von Neumann and many others who abandoned traditional
causality, were little help in resolving these paradoxes despite the
astonishing success in the prediction of quantum phenomena. Both the
implication and consequences of the quantum theory of light and matter, as
well as its profound mathematical, conceptual and philosophical foundations
are not yet understood completely by the majority of quantum physicists.

In order to appreciate the quantum drama which has been developing through
the whole century, and to estimate possible consequences of it in the new
quantum technological age, it seems useful to give a brief account of the
discovery of quantum theory at the beginning of the 20th century.

\subsection{The Discovery of Quantum Noise}

In 1918 Max Planck was awarded the Nobel Prize in Physics for his quantum
theory of blackbody radiation or as we would say now, quantum theory of
thermal noise based on the hypothesis of energy discontinuity. Invented in
1900, it inspired an unprecedented revolution in both physical science and
philosophy of the 20th century, with an unimaginable deep revision in our
way of thinking. As Planck stated later:-

\begin{quote}
\textit{If anyone says he can think about quantum problems without getting
giddy, that only shows he has not understood the first thing about them}
\end{quote}

\subsubsection{Quantum statistics for simpletons and children}

It is hard to realize that just 100 years ago the existence of the
fundamental quantum of action $\hbar $ had not been known. In autumn of 1900
Planck made two famous reports to Berlin Physics Society \cite{Plnkqp} about
the discovery of his constant. Although $\hbar $ evaluated by Planck is
very-very small, $\hbar \approx 10^{-34}\unit{kg}\unit{m}^{2}/\unit{s}$ when
measured in conventional units, such as kilograms, meters and seconds (and
in some places it is still being neglected), we can always take $\hbar =1$
by choosing suitable units.

There is a broad literature on the analysis Planck's 1900 paper, see for
example Landsberg 1981, \cite{Lan81}, for the analysis of his arguments from
the point of view of statistical thermodynamics and the literature cited in
this paper. Our analysis is much more elementary, it is simply based on
experimental data which had been known to Planck prior he derived his
formula.

According to the Boltzmann law of classical statistics, an absolutely black
body, which absorbs light of all colours, or frequencies, equally at a
temperature $\tau>0$, would radiate absolutely white light, consisting of
the mixture of all colours (frequencies $\omega$) of equal energies $E\left(
\tau,\omega\right) =k\tau$, where $k$ is the Boltzmann constant which can be
also set to one by choosing the suitable units of measurement.

Every child knows that it is not true: a heated blackbody (a piece of
burning coal) at lower temperatures radiates more red light, and becomes
white only at a high temperatures $\tau $. Planck's concern was to combine
the empirical formulas of Rayleigh ($E_{r}$) and Wien ($E_{w}$), 
\begin{equation*}
E_{r}\left( \omega \right) =k\tau -\frac{1}{2}\hbar \omega ,\quad \quad
\quad \quad E_{w}\left( \omega \right) =\hbar \omega e^{-\hbar \omega /k\tau
},
\end{equation*}

$\FRAME{itbpF}{2.0081in}{2.0081in}{0in}{}{}{}{\special{language "Scientific
Word";type "MAPLEPLOT";width 2.0081in;height 2.0081in;depth 0in;display
"PICT";plot_snapshots TRUE;function \TEXUX{$1-\frac{1}{2}\omega $};linecolor
"red";linestyle 1;linethickness 1;pointstyle "point";xmin "-5";xmax
"5";xviewmin "0";xviewmax "4";yviewmin "0";yviewmax
"1.2";viewset"XY";rangeset"X";phi 45;theta 45;plottype 4;plottickdisable
TRUE;labeloverrides 2;y-label "Ec";numpoints 49;axesstyle "normal";xis
\TEXUX{v969};var1name \TEXUX{$x$};valid_file "T";tempfilename
'G5D00601.wmf';tempfile-properties "XP";}}$\FRAME{itbpF}{2.5097in}{2.0081in}{%
0in}{}{}{}{\special{language "Scientific Word";type "MAPLEPLOT";width
2.5097in;height 2.0081in;depth 0in;display "PICT";plot_snapshots
TRUE;function \TEXUX{$\nu e^{-\nu }$};linecolor "red";linestyle
1;linethickness 1;pointstyle "point";xmin "-5";xmax "5";xviewmin
"0";xviewmax "5";yviewmin "0";yviewmax "1.2";viewset"XY";rangeset"X";phi
45;theta 45;plottype 4;plottickdisable TRUE;labeloverrides 3;x-label
"";y-label "Ew";numpoints 49;axesstyle "normal";xis \TEXUX{v957};var1name
\TEXUX{$x$};valid_file "T";tempfilename 'G5D00802.wmf';tempfile-properties
"XPR";}}\linebreak which approximate the spectral density of the energy
radiated by a blackbody at the temperature $\tau $ at a high and low
frequencies $\omega $ respectively. In October 1900 he announced the
Planck's quantum radiation formula $E\left( \tau ,\omega \right) =\hbar
\omega \left( e^{\hbar \omega /k\tau }-1\right) ^{-1}$ which becomes
classical $k\tau $ only at the limit $\omega \rightarrow \infty $. Thus the
constant (classical, $E_{c}$) spectral distribution of the thermal ``white
noise'' was replaced by the linear-exponential (quantum $E_{q}$) spectral
distribution: 
\begin{equation*}
E_{c}\left( \omega \right) =k\tau ,\quad \quad \quad \quad E_{q}\left(
\omega \right) =\hbar \omega \left( e^{\hbar \omega /k\tau }-1\right) ^{-1}
\end{equation*}

$\FRAME{itbpF}{2.0081in}{2.0081in}{0in}{}{}{}{\special{language "Scientific
Word";type "MAPLEPLOT";width 2.0081in;height 2.0081in;depth 0in;display
"PICT";plot_snapshots TRUE;function \TEXUX{$1$};linecolor "red";linestyle
1;linethickness 1;pointstyle "point";xmin "-5";xmax "5";xviewmin
"0";xviewmax "4";yviewmin "0";yviewmax "1.2";viewset"XY";rangeset"X";phi
45;theta 45;plottype 4;plottickdisable TRUE;labeloverrides 2;y-label
"Ec";numpoints 49;axesstyle "normal";xis \TEXUX{v57587};var1name
\TEXUX{$x$};valid_file "T";tempfilename 'G5D00A03.wmf';tempfile-properties
"XP";}}\FRAME{itbpF}{2.5097in}{2.0081in}{0.0104in}{}{}{}{\special{language
"Scientific Word";type "MAPLEPLOT";width 2.5097in;height 2.0081in;depth
0.0104in;display "PICT";plot_snapshots TRUE;function \TEXUX{$\nu \left(
e^{\nu }-1\right) ^{-1}$};linecolor "red";linestyle 1;linethickness
1;pointstyle "point";xmin "-5";xmax "5";xviewmin "0";xviewmax "5";yviewmin
"0";yviewmax "1.2";viewset"XY";rangeset"X";phi 45;theta 45;plottype
4;plottickdisable TRUE;labeloverrides 2;y-label "Eq";numpoints 49;axesstyle
"normal";xis \TEXUX{v957};var1name \TEXUX{$x$};valid_file "T";tempfilename
'G5D00B04.wmf';tempfile-properties "XP";}}$

\subsubsection{The A-level derivation of Planck's formula}

Within two months Planck made a complete theoretical deduction of his
formula, renouncing classical physics and introducing the weird assumption
that the radiated energy consist of discrete portions, or quanta which are
proportional to $\omega $: 
\begin{equation*}
\varepsilon _{n}\left( \omega \right) =\hbar \omega n,\quad n=0,1,2,\ldots .
\end{equation*}
By expanding $z/\left( 1-z\right) $ into the power series $\left( 1-z\right)
\sum nz^{n}$ with respect to $z=e^{-\hbar \omega /k\tau }$ he noted that the
right formula $E=E_{q}$ can be written as an averaging 
\begin{equation*}
\hbar \omega z\left( 1-z\right) ^{-1}=\sum_{n=0}^{\infty }\hbar \omega np_{n}
\end{equation*}
of the discrete energy levels $\varepsilon _{n}\left( \omega \right) $ with
the geometric probability distribution $p_{n}=\left( 1-z\right) z^{n}$. Thus
he concluded that 
\begin{equation*}
E\left( \tau ,\omega \right) =\hbar \omega N\left( \tau ,\omega \right) ,
\end{equation*}
where 
\begin{equation*}
N\left( \tau ,\omega \right) =\left( 1-e^{-\hbar \omega /k\tau }\right)
\sum_{n=0}^{\infty }ne^{-\hbar \omega n/k\tau }=\left( e^{\hbar \omega
/k\tau }-1\right) ^{-1}
\end{equation*}
is the average of the number of quanta for the Boltzmann probability
distribution 
\begin{equation*}
p_{n}=\left( 1-e^{-\hbar \omega /k\tau }\right) e^{-\hbar \omega n/k\tau }
\end{equation*}
of the discrete energy levels $\varepsilon _{n}\left( \omega \right) $. The
theory met resistance, and Planck himself ``tried at first to somehow weld
the elementary quantum of action somehow onto the framework of classical
theory. But in the face of all such attempts this constant showed itself to
be obdurate'' \cite{Plnkab}. He continued these futile attempts for a number
of years, and they cost him a great deal of efforts. He himself gave credit
to Boltzmann, as the only way to get the right spectral density was to
assume that it is the Boltzmann mean value of the discrete $\varepsilon _{n}$%
.

\subsubsection{The emergence of the quantum theory of light}

In 1905 Einstein, examining the photoelectric effect, proposed a quantum
theory of light, only later realizing that Planck's theory made implicit use
of this quantum light hypothesis. Einstein saw that the energy changes in a
quantum material oscillator occur in jumps which are multiples of $\omega$.
Einstein received Nobel prize in 1922 for his work on the photoelectric
effect. In 1924 Einstein arranged for the a publication of another important
paper by Bose, which had been initially rejected by a referee. Bose proposed
a new notion for statistical independence of quantum particles by putting
them into independent cells, and conjectured that there is no conservation
of the number of photons. Time has shown that Bose was right on all these
points.

Thus, quantum theory first emerged as the result of experimental data not in
the form of quantum mechanics but in the form of statistical observations of
quantum noise, the basic concept of quantum probability and quantum
stochastic processes. The corpuscular nature of light seemed to contradict
the Maxwell electromagnetic wave theory of light. In 1924 Einstein wrote:

\begin{quote}
\textit{There are therefore now two theories of light, both indispensable,
and - as one must admit today, despite twenty years of tremendous effort on
the part of theoretical physics - without any logical connection.}
\end{quote}

\subsection{The Rise of Quantum Mechanics}

The solution to the paradox of the wave/corpuscular duality of light came
unexpectedly when de Broglie made the even more bizarre conjecture of
extending this duality to matter. As a young girl said later to Schr\"{o}%
dinger, who discovered the quantum mechanics of wave matter:

\begin{quote}
\textit{Hey, you never even thought when you began that so much sensible
stuff would come out of it}
\end{quote}

\textit{. } (quoted from the Preface in \cite{Schr26c}).

\subsubsection{The discovery of matrix mechanics}

In 1912 Niels Bohr worked in the Rutherford group in Manchester on his
theory of the electron in an atom. He was puzzled by the discrete spectra of
light which is emitted by atoms when they are subjected to an excitation. He
was influenced by the ideas of Planck and Einstein and addressed a certain
paradox in his work. How can energy be conserved when some energy changes
are continuous and some are discontinuous, i.e. change by quantum amounts?%
\textit{\ }Bohr conjectured that an atom could exist only in a discrete set
of stable energy states, the differences of which amount to the observed
energy quanta. Bohr returned to Copenhagen and published a revolutionary
paper on the hydrogen atom in the next year. He suggested his famous formula 
\begin{equation*}
E_{m}-E_{n}=\hbar\omega_{mn}
\end{equation*}
from which he derived the major laws which describe physically observed
spectral lines. This work earned Niels Bohr the 1922 Nobel Prize about $%
10^{5}$ Swedish Kroner.

In 1925 a young German theoretical physicist, Heisenberg, gave a preliminary
account of a new and highly original approach to the mechanics of the atom 
\cite{Heis25}. He was influenced by Niels Bohr and proposed to substitute
for the position coordinate of an electron in the atom arrays 
\begin{equation*}
q_{mn}\left( t\right) =q_{mn}e^{i\omega_{mn}t}
\end{equation*}
indexed by Bohr's differences $E_{m}-E_{n}$. They would account for the
random jumps $E_{m}\mapsto E_{n}$ in the atom corresponding to the
spontaneous emission of the energy quanta $\hbar\omega_{mn}$ for the
Planck's electro-magnetic oscillators 
\begin{equation*}
\frac{\mathrm{d}}{\mathrm{d}t}q_{mn}\left( t\right)
=i\omega_{mn}q_{mn}\left( t\right) .
\end{equation*}
His Professor, Max Born, was a mathematician who immediately recognized an
infinite matrix algebra in Heisenberg's multiplication rule for the tables $%
Q=\left[ q_{mn}\right] $. The classical momentum was also replaced by a
similar matrix, 
\begin{equation*}
P\left( t\right) =\left[ p_{mn}e^{i\omega_{mn}t}\right]
\end{equation*}
and $P$ and $Q$ matrices were postulated to follow a commutation law: 
\begin{equation*}
\left[ Q\left( t\right) ,P\left( t\right) \right] =i\hbar I,
\end{equation*}
where $I$ is the unit matrix. The classical Hamiltonian equations of
dynamical evolution were now replaced by 
\begin{equation*}
\frac{\mathrm{d}}{\mathrm{d}t}Q\left( t\right) =\frac{i}{\hbar}\left[
H,Q\left( t\right) \right] ,\;\frac{\mathrm{d}}{\mathrm{d}t}P\left( t\right)
=\frac{i}{\hbar}\left[ H,P\left( t\right) \right] ,
\end{equation*}
where $H=\left[ E_{n}\delta_{mn}\right] $ is the diagonal Hamilton matrix 
\cite{BHJ26}. Thus quantum mechanics was first invented in the form of 
\textit{matrix mechanics}, emphasizing the possibilities of quantum
transitions, or jumps between the stable energy states $E_{n}$ of an
electron. In 1932 Heisenberg was awarded the Nobel Prize for his work in
mathematical physics.

Conceptually, the new atomic theory was based on the positivism of Mach as
it operated not with real space-time but with only observable quantities
like atomic transitions. However many leading physicists were greatly
troubled by the prospect of loosing reality and deterministic causality in
the emerging quantum physics. Einstein, in particular, worried about the
element of `chance' which had entered physics. In fact, this worries came
rather late since Rutherford had introduced a spontaneous effect when
discussing radio-active decay in 1900.

\subsubsection{The discovery of wave mechanics}

In 1923 de Broglie, inspired by the works of Einstein and Planck, extended
the wave-corpuscular duality also to material particles. He used the
Hamilton-Jacobi theory which had been applied both to particles and waves.
In 1928 de Broglie received the Nobel Prize for this work.

In 1925, Schr\"{o}dinger gave a seminar on de Broglie's material waves, and
a member of the audience suggested that there should be a wave equation.
Within a few weeks Schr\"{o}dinger found his celebrated wave equation, first
in a relativistic, and then in the non-relativistic form \cite{Schr26}.
Instead of seeking the classical solutions to the Hamilton-Jacobi equation 
\begin{equation*}
H\left( q,\frac{\hbar}{i}\frac{\partial}{\partial q}\ln\psi\right) =E
\end{equation*}
he suggested finding those wave functions $\psi$ which satisfy the linear
equation 
\begin{equation*}
H\left( q,\frac{\hbar}{i}\frac{\partial}{\partial q}\right) \psi=E\psi
\end{equation*}
(It coincides with the former equation only if $H$ is linear with respect to 
$p$).

Schr\"{o}dinger published his revolutionary \textit{wave mechanics} in a
series of six papers \cite{Schr26c} in 1926 during a short period of
sustained creative activity that is without parallel in the history of
science. Like Shakespeare, whose sonnets were inspired by a dark lady, Schr%
\"{o}dinger was inspired by a mysterious lady of Arosa where he took ski
holidays during the Christmas 1925 but `had been distracted by a few
calculations'. This was the second formulation of quantum theory, which he
successfully applied to the Hydrogen atom, oscillator and other quantum
mechanical systems, solving the corresponding Sturm-Liouville problems. The
mathematical equivalence between the two formulations of quantum mechanics
was understood by Schr\"{o}dinger in the fourth paper where he obtained the
non-stationary wave equation written in terms of the Hamiltonian operator $%
H=H\left( q,\frac{\hbar }{i}\frac{\partial }{\partial q}\right) $ as 
\begin{equation*}
i\hbar \frac{\partial }{\partial t}\psi \left( t\right) =H\psi \left(
t\right) ,
\end{equation*}
and he also introduced operators associated with each dynamical variable.

Unlike Heisenberg and Born's matrix mechanics, the general reaction towards
wave mechanics was immediately enthusiastic. Plank described Schr\"{o}%
dinger's wave mechanics as `epoch-making work'. Einstein wrote: `the idea of
your work springs from true genius...'. Next year Schr\"{o}dinger was
nominated for the Nobel Prize, but he failed to receive it in this and five
further consecutive years of his nominations by most distinguished
physicists of the world, the reason behind his rejection being `the highly
mathematical character of his work'. Only in 1933 did he receive his prize,
this time jointly with Dirac, and this was the first, and perhaps the last,
time when the Nobel Prize for physics was given to true mathematical
physicists.

Following de Broglie, Schr\"{o}dinger initially thought that the wave
function corresponds to a physical vibration process in a real continuous
space-time because it was not stochastic, but he was puzzled by the failure
to explain the blackbody radiation and photoelectric effect from this wave
point of view. In fact the wave interpretation applied to light quanta leads
back to classical electrodynamics, as his relativistic wave equation for a
single photon coincides mathematically with the classical wave equation.
However after realizing that the time-dependent $\psi$ is a complex function
in his fourth 1926 paper \cite{Schr26c}, he admitted that the wave function $%
\psi$ cannot be given a direct interpretation, and described the wave
density $\psi\bar{\psi}$ as a sort of weight function for superposition of
point-mechanical configurations.

Although Schr\"{o}dinger was a champion of the idea that the most
fundamental laws of the microscopic world are absolutely random even before
he discovered wave mechanics, he failed to see the probabilistic nature of $%
\psi \bar{\psi}$. Indeed, his equation was not stochastic, and it didn't
account for the random jumps $E_{m}\rightarrow E_{n}$ of the Bohr-Heisenberg
theory but rather opposite, it did prescribe the preservation of the
eigenvalues $E=E_{n}$. He even wrote:

\begin{quote}
\textit{If we have to go on with these dammed quantum jumps, then I'm sorry
that I ever got involved.}
\end{quote}

For the rest of his life Schr\"{o}dinger was trying to find from time to
time without a success a more fundamental equation which would be
responsible for the energy transitions in the process of measurement of the
quanta $\hbar\omega_{mn}$. As we shall see, he was right assuming the
existence of such equation.

\subsubsection{Interpretations of quantum mechanics}

The creators of the rival matrix quantum mechanics were forced to accept the
simplicity and beauty of Schr\"{o}dinger's approach. In 1926 Max Born put
forward the statistical interpretation of the wave function by introducing
the statistical mean 
\begin{equation*}
\left\langle H\right\rangle =\int \bar{\psi}\left( x\right) H\psi \left(
x\right) \mathrm{d}x
\end{equation*}
for a dynamical variable $H$ in the physical state, described by $\psi $.
This was developed in Copenhagen and gradually was accepted by almost all
physicists as the ``Copenhagen interpretation''. Born by education was a
mathematician, and he would be the only mathematician ever to receive the
Nobel Prize (in 1953, for his statistical studies of wave functions) were it
not for the fact that he became a physicist, later Professor of Natural
Philosophy at Edinburgh. Bohr, Born and Heisenberg considered electrons and
quanta as unpredictable particles which cannot be visualized in the real
space and time.

The most outspoken opponent of a/the probabilistic interpretation was
Einstein. Albert Einstein admired the new development of quantum theory but
was suspicious, rejecting its acausality and probabilistic interpretation.
It was against his scientific instinct to accept statistical interpretation
of quantum mechanics as a complete description of physical reality. There
are famous sayings of his on that account:

\begin{quote}
\textit{`God is subtle but he is not malicious', `God doesn't play dice'}
\end{quote}

During these debates on the probabilistic interpretation of quantum
mechanics of Einstein between Niels Bohr, Schr\"{o}dinger often sided with
his friend Einstein, and this may explain why he was distancing himself from
the statistical interpretation of his wave function. Bohr invited Schr\"{o}%
dinger to Copenhagen and tried to convince him of the particle-probabilistic
interpretation of quantum mechanics. The discussion between them went on day
and night, without reaching any agreement. The conversation, however deeply
affected both men. Schr\"{o}dinger recognized the necessity of admitting
both wave and particles, but he never devised a comprehensive interpretation
rival to Copenhagen orthodoxy. \ Bohr ventured more deeply into
philosophical waters and emerged with his concept of complementarity:

\begin{quote}
\textit{Evidence obtained under different experimental conditions cannot be
comprehended within a single picture, but must be regarded as complementary
in the sense that only the totality of the phenomena exhausts the possible
information about the objects.}
\end{quote}

In his later papers Schr\"{o}dinger accepted the probabilistic
interpretation of $\psi \bar{\psi}$, but he did not consider these
probabilities classically, instead he interpreted them as the strength of
our belief or anticipation of an experimental result. In this sense the
probabilities are closer to propensities than to the frequencies of the
statistical interpretation of Born and Heisenberg. Schr\"{o}dinger had never
accepted the subjective positivism of Bohr and Heisenberg, and his
philosophy is closer to that called representational realism. He was content
to remain a critical unbeliever.

There have been many other attempts to retain the deterministic realism in
the quantum world, the most extravagant among these being the ensemble-world
interpretations of Bohm \cite{Boh52} and Everett \cite{Eve57}. The first
interpretational theory, known as the pilot-wave theory, is based on the
conventional Schr\"{o}dinger equation which is used to define the flow of a
classical fluid in the configuration space. The predictions of this
classical macroscopic theory coincide with the statistical predictions of
the orthodox quantum theory only for the ensembles of coordinate-like
observables with the initial probability distribution over the many worlds
given by the initial pilot wave. Other observables like momenta which are
precisely determined at each point by the velocity of this fluid, have no
uncertainty under the fixed coordinates. This is inconsistent with the
prediction of quantum theory for individual systems, and there is no way to
incorporate the stochastic dynamics of sequentially monitored individual
quantum particles in the single world into this fluid dynamics. Certainly
this is a variation of the de Broglie-Schr\"{o}dinger old interpretation,
and it doesn't respect the Bell's first principle for the interpretational
theories ``that it should be possible to formulate them for small systems'' 
\cite{Bell87}, p. 126.

The Everett's many-world interpretation also assumes that the classical
configurations at each time are distributed in the comparison class of
possible worlds worth probability density $\psi\bar{\psi}$. However no
continuity between present and past configurations is assumed, and all
possible outcomes of their measurement are realized every time, each in a
different edition of the continuously multiplying universe. The observer in
a given brunch of the universe is aware only of what is going on in that
particular branch, and this results in the reduction of the wave-function.
This would be macroscopically equivalent to the pilot-wave theory if the de
Broglie-Schr\"{o}dinger-Bohm fluid dynamics could be obtained as the average
of wave equations over all brunches. Every statistician experienced in the
classical continuous measurements would recognize in this continuously
branching model the many-world re-interpretation for a stochastic Markov
process, and would apply the well-developed stochastic differential calculus
to analyze this dynamical model. However no such equation for a continuously
monitored branch was suggested in this theory, although there should be many
if at all, corresponding to many possible choices of classical
configurations (e.g. positions or momenta) for a single many-world Schr\"{o}%
dinger equation.

Certainly there are some advantages living in many worlds from the
philosophical point of view, from the practical point of view, however, to
have an infinite number (continuum product of continua?) of real worlds at
the same time seems not better than to have none. As Bell wrote in \cite%
{Bell87}, p. 134: ``to have multiple universes, to realize all possible
configurations of particles, would have seemed grotesque''. Even if such a
weighted many-world dynamical theory had been developed to a satisfactory
level, it would be immediately reformulated as a stochastic evolutionary
theory in our single world with well-established mathematical language and
statistical interpretation. In fact, the stochastic theory of continuously
observed quantum systems has been already derived, not just developed, in
full generality and rigor in quantum probability, and it will be presented
in the last section. But first I shall demonstrate the underlying ideas on
the elementary single-transition level.

\section{Quantum Uncertainties and Paradoxes}

\medskip

\begin{quote}
\textit{How wonderful we have met with a paradox, now we have some hope of
making progress} - Niels Bohr.
\end{quote}

In 1932 von Neumann put quantum theory on firm theoretical basis by setting
the mathematical foundation for new, quantum, probability theory, the
quantitative theory for counting non commuting events of quantum logics.
This noncommutative probability theory is based on essentially more general
axioms than the classical (Kolmogorovian) probability of commuting events,
which form common sense Boolean logic, first formalized by Aristotle. It has
been under extensive development during the last 30 years since the
introduction of algebraic and operational approaches for treatment of
noncommutative probabilities, and currently serves as the mathematical basis
for quantum information and measurement theory.

Here we shall demonstrate the main ideas of quantum probability and quantum
paradoxes arising from the application of classical probability theory to
quantum phenomena on the simple examples of a single quantum system such as
quantum bit (qubit, or q-bit). One can identify it with a single spin $1/2$
in the famous EPR paradox, or with an unstable atom in a single state as in
the Schr\"{o}dinger's quantum measurement model with his cat. The most
recent mathematical development of these models and methods leads to a
profound quantum filtering and control theory in quantum open systems
presented in the Section 3 which has found numerous applications in quantum
statistics, optics and spectroscopy, and is an appropriate tool for the
solution of the dynamical decoherence problem for quantum communications and
computations.

\subsection{Uncertainties and Quantum Logics}

Bohr was concerned with the paradox of spontaneous emission. He addressed
the question: How does the electron know when to emit radiation? Bohr, Born
and Heisenberg abandoned causality of traditional physics in the most
positivistic way. Max Born said:

\begin{quote}
\textit{If God made the world a perfect mechanism, ... we need not solve
innumerable differential equations, but can use dice with fair success.}
\end{quote}

\subsubsection{Heisenberg uncertainty relations}

In 1927 Heisenberg derived \cite{Heis27} his famous uncertainty relation 
\begin{equation*}
\Delta Q\Delta P\geq\hbar/2,\quad\Delta T\Delta E\geq\hbar/2
\end{equation*}
which gave mathematical support to the revolutionary complementary principle
of Bohr. The first relation was easily proved in the Schr\"{o}dinger
representations $Q=x$, $P=\frac{\hbar}{i}\frac{\partial}{\partial x}$ in
terms of the standard deviations 
\begin{equation*}
\Delta Q=\left( \left\langle Q^{2}\right\rangle -\left\langle Q\right\rangle
^{2}\right) ^{1/2},\quad\Delta P=\left( \left\langle P^{2}\right\rangle
-\left\langle P\right\rangle ^{2}\right) ^{1/2}.
\end{equation*}
The second relation, which was first stated by analogy of $t$ with $x$ and
of $E$ with $P$, can be proved \cite{Be76, Hol80} in terms of the optimal
measurement of the initial time as an unknown parameter $\tau$ of the Schr%
\"{o}dinger's state $\psi\left( t-\tau\right) $ which is realized by the
measurement of self-adjoint operator $T=t$ in an extended representation
where the Hamiltonian $H$ is given by the operator $E=i\hbar\frac{\partial}{%
\partial t}$. As Dirac stated:

\begin{quote}
\textit{Now when Heisenberg noticed that, he was really scared.}
\end{quote}

Einstein launched an attack on the uncertainty relation at the Solvay
Congress in 1927, and then again in 1930, by proposing cleverly devised
thought experiments which would violate this relation. Most of these
imaginary experiments were designed to show that interaction between the
microphysical object and the measuring instrument is not so inscrutable as
Heisenberg and Bohr maintained. He suggested, for example, a box filled with
radiation with a clock. The clock is designed to open a shutter and allow
one photon to escape. By weighing the box the photon energy and the time of
escape can both be measured with arbitrary accuracy.

After proposing this argument Einstein is reported to have spent a happy
evening, and Niels Bohr an unhappy one. After a sleepless night he showed
next morning that Einstein was wrong. Mathematically his solution can be
expressed by the following formula of `signal plus noise' 
\begin{equation*}
X=t+Q,\quad Y=i\hbar \frac{\partial }{\partial t}+P
\end{equation*}
for the measuring quantity $X$, the pointer coordinate of the clock, and the
observable $Y$ for indirect measurement of photon energy $E=i\hbar \frac{%
\partial }{\partial t}$ in the Einstein experiment, where $Q$ and $P$ are
the position and momentum operators of the compensation weight under the
box. Due to the initial independence of the weight, the commuting
observables $X$ and $Y$ have even greater uncertainty 
\begin{equation*}
\Delta X\Delta Y=\Delta T\Delta E+\Delta Q\Delta P\geq \hbar
\end{equation*}
than that predicted by Heisenberg uncertainty $\Delta T\Delta E\geq \hbar /2$%
.

\subsubsection{Nonexistence of hidden variables}

Einstein hoped that eventually it would be possible to explain the
uncertainty relations by expressing quantum mechanical observables as
functions of some hidden variables $\lambda $ in deterministic physical
states such that the statistical aspect will arise as in classical
statistical mechanics by averaging these observables over $\lambda $.

Von Neumann's monumental book \cite{Neum32} on the mathematical foundations
of quantum theory was therefore a timely contribution, clarifying, as it
did, this point. \ Inspired by Lev Landau, he introduced, for the unique
characterization of the statistics of a quantum ensemble, the statistical
density operator $\rho $ which eventually, under the name normal, or regular
state, became a major tool in quantum statistics. He considered the linear
space $\mathcal{L}$ of all bounded Hermitian operators $L=L^{\ast }$ as
potential obsrvables in a quantum system described by a Hilbert space $%
\mathbb{H}$ of all (not yet normalized to one) wave functions $\psi $.
Although von Neumann considered any complete inner product complex linear
space as the Hilbert space, it is sufficient to reproduce his analysis for a
finite-dimensional $\mathbb{H}$. He defined the expectation $\left\langle
L\right\rangle $ of each $L\in \mathcal{L}$ in a state $\rho $ by the
continuous (ultra-weakly continuous if $\mathbb{H}$ is infinite-dimensional)
linear functional $\left\langle L\right\rangle =\mathrm{Tr}L\rho $, where $%
\mathrm{Tr}$ denotes the linear operation of trace applied to the product of
all operators on the right. He noted that in order to have positive
probabilities for quantum mechanical events $E=1$ as expectations $%
\left\langle E\right\rangle $ of yes-no observables $E\in \mathcal{L}$ with $%
\left\{ 0,1\right\} $ spectrum, and probability one for the identity event $%
I=1$ described by identity operator $I$ (the multiplication by $1$), 
\begin{equation*}
\Pr \left\{ E=1\right\} =\mathrm{Tr}E\rho \geq 0,\quad \Pr \left\{
I=1\right\} =\mathrm{Tr}\rho =1,
\end{equation*}
the statistical operator $\rho $ must be positive-definite and have trace
one. Then he proved that any\emph{\ physically continuous additive}
functional $L\mapsto \left\langle L\right\rangle $ is regular, i.e. has such
trace form. He applied this technique to the analysis of the completeness
problem of quantum theory, i.e. whether it constitutes a logically closed
theory or whether it could be reformulated as an entirely deterministic
theory through the introduction of hidden parameters (additional variables
which, unlike ordinary observables, are inaccessible to measurements). He
came to the conclusion that

\begin{quote}
\textit{the present system of quantum mechanics would have to be objectively
false, in order that another description of the elementary process than the
statistical one may be possible}

(quoted on page 325 in \cite{Neum32})
\end{quote}

To prove this theorem, von Neumann showed that there is no such state which
would be dispersion-free simultaneously for all quantum events $E\in 
\mathcal{L}$ described by Hermitian projectors $E^{2}=E$. For each such
state, he argued, 
\begin{equation*}
\left\langle E^{2}\right\rangle =\left\langle E\right\rangle =\left\langle
E\right\rangle ^{2}
\end{equation*}
for all events $E$ would imply that $\rho =O$, which cannot be statistical
operator as $\mathrm{Tr}O=0\neq 1$. Thus no state can be considered as a
mixture of dispersion-free states, each of them associated with a definite
value of hidden parameters. There are simply no such states, and thus, no
hidden parameters. In particular this implies that the statistical nature of
pure states, which are described by one-dimensional projectors $\rho
=P_{\psi }$ corresponding to wave functions $\psi $, cannot be removed by
supposing them to be a mixture of dispersion-free substates.

It is widely believed that in 1966 John Bell showed that von Neuman's proof
was in error, or at least his analysis left the real question untouched \cite%
{Bell66}. To discredit the von Neumann's proof he constructed an example of
dispersion-free states parametrized for each quantum state $\rho $ by a real
parameter $\lambda $ for a single quantum bit corresponding to the two
dimensional $\mathbb{H=}\mathfrak{h}$ (we use the little $\mathfrak{h}\simeq 
\mathbb{C}^{2}$ in order to emphasize that this is the simplest quantum
system). He succeeded to do this by weakening the assumption of the
additivity for such states, requiring it only for the commuting observables
in $\mathcal{L}$, and by abandoning the linearity of the constructed
expectations in $\rho $ described by spin polarization vector $\mathbf{r}$.
There is no reason, he argued, to keep the linearity in $\rho $ for the
observable eigenvalues determined by $\lambda $ and $\rho $, and to demand
the additivity for non-commuting observables as they are not simultaneously
measurable: The measured eigenvalues of a sum of noncommuting observables
are not the sums of the eigenvalues of this observables measured separately.
For each spin-projection $L=\sigma \left( \mathbf{l}\right) $ given by a
3-vector $\mathbf{l}$ Bell found a family $s_{\lambda }\left( \mathbf{l}%
\right) $ of dispersion-free values $\pm l$, $l=\left| \mathbf{l}\right| $,
parameterized by $\left| \lambda \right| \leq 1/2$, which reproduce the
expectation $\left\langle \sigma \left( \mathbf{l}\right) \right\rangle =%
\mathbf{l\cdot r} $ in the pure quantum state when uniformly averaged over
the $\lambda $. However his example does not contradict the von Neumann
theorem even if the latter is strengthened by the restriction of the
additivity only to the commuting observables: The constructed
dispersion-free expectation function $L\mapsto \left\langle L\right\rangle
_{\lambda }$ is not \emph{physically continuous} on $\mathcal{L}$ because
the value $\left\langle L\right\rangle _{\lambda }=s_{\lambda }\left( 
\mathbf{l}\right) $ is one of the eigenvalues $\pm 1$ for each $\lambda $,
and it covers both values when the directional vector $\mathbf{l}$ rotates
continuously over the three-dimensional sphere. A function $\mathbf{l\mapsto 
}\left\langle \sigma \left( \mathbf{l}\right) \right\rangle _{\lambda }$ on
the continuous manifold (sphere) with discontinuous values can be continuous
only if it is constant, but this is ruled out by the demand to reproduce the
expectation $\left\langle \sigma \left( \mathbf{l}\right) \right\rangle =%
\mathbf{l\cdot r}$ which is variable in $\mathbf{l}$ by averaging over $%
\lambda $ the constants $\left\langle \sigma \left( \mathbf{l}\right)
\right\rangle _{\lambda }$ in $\mathbf{l}$. Measurements of the projections
of spin on the physically close directions should be described by close
expected values in any physical state specified by $\lambda $, otherwise it
cannot have physical meaning! More detailed critical analysis of the Bell's
arguments is given in the Appendix 1$.$

Since then there were innumerable attempts to introduce hidden variables in
ever more sophisticated forms, perhaps not yet discovered, which would
determine the complementary variables if the hidden variables were measured
precisely. In higher dimensions of $\mathbb{H}$ all these attempts are ruled
out by Gleason's theorem \cite{Glea57} who proved that there is no even one
additive zero-one probability function if $\dim \mathbb{H}>2$.

\subsubsection{Complementarity and common sense}

In view of the decisive importance of this analysis for the foundations of
quantum theory, Birkhoff and von Neumann \cite{BiNe36} setup a system of
formal axioms for the calculus of logicotheoretical propositions concerning
results of possible measurements in a quantum system. They started by
formalizing the calculus of \emph{quantum propositions} $E\in \mathcal{E}$
corresponding to the idealized events described by orthoprojectors $E$ on a
Hilbert space $\mathbb{H}$, the projective operators $E=E^{2}$ which are
orthogonal to their complements $E^{\bot }=I-E$ in the sense $E^{\ast
}E^{\bot }=O$, where $O$ denotes the multiplication by $0.$ This is
equivalent to 
\begin{equation*}
E^{\ast }=E^{\ast }E=E,
\end{equation*}
i.e. the set of propositions $\mathcal{E}$ is the set of all Hermitian
projectors $E\in \mathcal{L}$ which are the only observables with two
eigenvalues $\left\{ 1,0\right\} $ (``yes'' and ``no''). Such calculus
coincides with the calculus of linear subspaces $\mathfrak{e}\subseteq 
\mathbb{H} $ including $0$-dimensional subspace $\mathbb{O}$, in the same
sense as the common sense propositional calculus of classical events
coincides with the calculus in a Boolean algebra of subsets $E\subseteq
\Omega $ including empty subset $\emptyset $. The subspaces $\mathfrak{e}$
are defined by the condition $\mathfrak{e}^{\bot \bot }=\mathfrak{e}$, where 
$\mathfrak{e}^{\bot }$ denotes the orthogonal complement $\left\{ \chi \in 
\mathbb{H}:\left\langle \chi |\psi \right\rangle =0,\psi \in \mathfrak{e}%
\right\} $ of $\mathfrak{e}$, and they uniquely define the propositions $E$
as the orthoprojectors $P\left( \mathfrak{e}\right) $ onto the ranges 
\begin{equation*}
\mathfrak{e}=\mathrm{range}E:=E\mathbb{H}
\end{equation*}
of $E\in \mathcal{E}$.\ In this calculus the logical ordering $E\leq F$
implemented by the algebraic relation $EF=E$ coincides with 
\begin{equation*}
\mathrm{range}E\subseteq \mathrm{range}F,
\end{equation*}
the conjunction $E\wedge F$ corresponds to the intersection, 
\begin{equation*}
\mathrm{range}\left( E\wedge F\right) =\mathrm{range}E\cap \mathrm{range}F,
\end{equation*}
however the disjunction $E\vee F$ is represented by the linear sum $%
\mathfrak{e}+\mathfrak{f}$ of the corresponding subspaces but not their
union 
\begin{equation*}
\mathrm{range}E\cup \mathrm{range}F\subseteq \mathrm{range}\left( E\vee
F\right) ,
\end{equation*}
and the smallest orthoprojector $O$ corresponds to zero-dimensional subspace 
$\mathbb{O=}\left\{ 0\right\} $ but not the empty subset $\emptyset $ (which
is not linear subspace). Note that although $\mathrm{range}\left( E+F\right)
=\mathfrak{e}+\mathfrak{f}$ $\ $for any $E,F\in \mathcal{E}$, the operator $%
E+F$ is not the orthoprojector $E\vee F$ corresponding to $\mathfrak{e}+%
\mathfrak{f}$ unless $EF=0$. This implies that the distributive law
characteristic for propositional calculus of classical logics no longer
holds. However it still holds for compatible propositions described by
commutative orthoprojectors due to the orthomodularity property 
\begin{equation*}
E\leq I-F\leq G\Longrightarrow \left( E\vee F\right) \wedge G=E\vee \left(
F\wedge G\right) .
\end{equation*}

Given the regular state $\left\langle E\right\rangle =\mathrm{Tr}E\rho $ on $%
\mathcal{E}$ one can\ also inroduce the probability measure 
\begin{equation*}
\mathrm{P}\left( \mathfrak{e}\right) =\Pr \left\{ P\left( \mathfrak{e}%
\right) =1\right\} =\left\langle P\left( \mathfrak{e}\right) \right\rangle
\end{equation*}
which is additive but only for orthogonal $\mathfrak{e}$ and $\mathfrak{f}$: 
\begin{equation*}
\mathfrak{e}\perp \mathfrak{f}\Rightarrow \mathrm{P}\left( \mathfrak{e}+%
\mathfrak{f}\right) =\mathrm{P}\left( \mathfrak{e}\right) +\mathrm{P}\left( 
\mathfrak{f}\right) .
\end{equation*}

Two propositions $E,F$ are called complementary if $E\vee F=I$,
orthocomplementary if $E+F=I$, incompatible or disjunctive if $E\wedge F=0$,
and contradictory or orthogonal if $EF=0$. As in the classical, common sense
case, logic contradictory propositions are incompatible. However \emph{%
incompatible propositions are not necessary contradictory }as can be easily
seen for any two nonorthogonal but not coinciding one-dimensional subspaces.
In particular, in quantum logics there exist complementary incompatible
pairs $E,F$, $E\vee F=I$, $E\wedge F=0$ which are not ortho-complementary in
the sense $E+F\neq I$, i.e. $EF\neq 0$ (this would be impossible in the
classical case). This is a rigorous logico-mathematical proof of Bohr's
complementarity.

As an example, we can consider the proposition that a quantum system is in a
stable energy state $E$, and an incompatible proposition $F$, that it
collapses at a given time $t$, say. The incompatibility $E\wedge F=0$
follows from the fact that there is no state in which the system would
collapse preserving its energy, however these two propositions are not
contradictory (i.e. not orthogonal, $EF\neq 0$): the system might not
collapse if it is in other than $E$ stationary state (remember the Schr\"{o}%
dinger's earlier belief that the energy law is valid only on average, and is
violated in the process of radiation).

In 1952 Wick, Wightman, and Wigner \cite{WWW52} showed that there are
physical systems for which not every orthoprojector corresponds to an
observable event, so that not every one-dimensional orthoprojector $\rho
=P_{\psi }$ corresponding to a wave function $\psi $ is a pure state. This
is equivalent to the admission of some selective events which are
dispersion-free in all pure states. Jauch and Piron \cite{JaPi63}
incorporated this situation into quantum logics and proved in the context of
this most general approach that the hidden variable interpretation is only
possible if the theory is observably wrong, i.e. if incompatible events are
in fact compatible or contradictory.

Bell criticized this as well as the Gleason's theorem, but this time his
arguments were not based even on the classical ground of usual probability
theory. Although he explicitly used the additivity of the probability on the
orthogonal events in his counterexample for $\mathbb{H}=\mathbb{C}^{2}$, he
questioned : `That so much follows from such apparently innocent assumptions
leads us to question their innocence'. (p.8 in \cite{Bell87}). In fact this
was equivalent to questioning the additivity of classical probability on the
corresponding disjoint subsets, but he didn't suggest any other complete
system of physically reasonable axioms for introducing such peculiar
``nonclassical'' hidden variables, not even a single counterexample to the
orthogonal nonadditivity for the simplest case of quantum bit $\mathbb{H}=%
\mathfrak{h}$. Thus Bell implicitly rejected classical probability theory in
the quantum world, but he didn't want to accept quantum probability as the
only possible theory for explaining the microworld. Even if such attempt was
successful for a single quantum system (as he possibly thought his
unphysical discontinuous construction in the case $\dim \mathbb{H}=2$ was),
it would satisfy only the classical composition law. This would not allow
extension of the dispersion-free product-states to a composed quantum system
because of their nonadditivity on the space $\mathcal{L}$ of all composed
observables in the Hilbert space $\mathbb{H}\otimes \mathfrak{g}$ with $%
\mathfrak{g}=\mathbb{C}^{n}$ for $n>1$. As it is shown in the Appendix 1,
the quantum composition law together with the orthoadditivity excludes the
hidden variable possibility also for $\mathbb{H}=\mathbb{C}^{2}$. Otherwise
it would contradict Gleason's theorem because $\dim \left( \mathfrak{h}%
\otimes \mathfrak{g}\right) =2n>2$, not to mention a hidden variable
reproduction of nonseparable, entangled states. Thus the quantum composition
principle justifies the von Neumann's additivity assumption for the states
on the whole operator algebra $\mathcal{B}\left( \mathbb{H}\right) $ of each
Hilbert space $\mathbb{H}$.

\subsection{Entanglement of Quantum Bits}

Heisenberg derived from the uncertainty relation that `the nonvalidity of
rigorous causality is necessary and not just consistently possible'. Max
Born even stated:

\begin{quote}
\textit{One does not get an answer to the question, what is the state after
collision? but only to the question, how probable is a given effect of the
collision?}
\end{quote}

The general scientific consensus in the physical world is that no positive
solution exists to these negative statements at present. And this will be so
unless we formulate these problems in a rigorous way and disagree with the
notorious saying \cite{Russ} that\ in mathematics ``we never know what we
are talking about'' ( B. Russel should have better said that ``we never know
what \emph{mathematicians} are talking about'': he was not a true
mathematician, the mathematicians know precisely what they are talking
about).

\subsubsection{Spooky action at distance}

After his defeat on uncertainty relations Einstein seemed to have become
resigned to the statistical interpretation of quantum theory, and at the
1933 Solvay Congress he listened to Bohr's paper on completeness of quantum
theory without objections. Then, in 1935, he launched a brilliant and subtle
new attack in a paper \cite{EPR} with two young co-authors, Podolski and
Rosen, which is known as the EPR paradox that has become of major importance
to the world view of physics. They stated the following requirement for a
complete theory as a seemingly necessary one:

\begin{quote}
\textit{Every element of physical reality must have a counterpart in the
physical theory.}
\end{quote}

The question of completeness is thus easily answered as soon as soon as we
are able to decide what are the elements of the physical reality. \ EPR then
proposed a sufficient condition for an element of physical reality:

\begin{quote}
\textit{If, without in any way disturbing the system, we can predict with
certainty the value of a physical quantity, then there exists an element of
physical reality corresponding to this quantity.}
\end{quote}

Then they designed a thought experiment the essence of which is that two
quantum ``bits'', particle spins of two electrons say, are brought together
to interact, and after separation an experiment is made to measure the spin
orientation of one of them. The state after interaction is such that the
measurement result $\tau =\pm \frac{1}{2}$ of one particle uniquely
determination the spin $z$-orientation $\sigma =\mp \frac{1}{2}$ of the
other particle. EPR apply their criterion of local reality: since the value
of $\sigma $ can be predicted by measuring $\tau $ without in any way
disturbing $\sigma $, it must correspond to an existing element of physical
reality. Yet the conclusion contradicts a fundamental postulate of quantum
mechanics, according to which the sign of spin is not an intrinsic property
of a complete description of the spin but is evoked only by a process of
measurement. Therefore, EPR conclude, quantum mechanics must be incomplete,
there must be hidden variables not yet discovered, which determine the spin
as an intrinsic property. It seems Einstein was unaware of the von Neumann's
theorem, although they both had positions at the Institute for Advanced
Studies at Princeton (being among the original six mathematics professors
appointed there in 1933).

Bohr carefully replied to this challenge by rejecting the assumption of
local physical realism as stated by EPR \cite{Bohr35}: `There is no question
of a mechanical disturbance of the system under investigation during the
last critical stage of the measuring procedure. But even at this stage there
is essentially a question of \textit{an influence on the very conditions
which define the possible types of predictions regarding the future behavior
of the system}'. This influence became notoriously famous as Bohr's \textit{%
spooky action at a distance}. He had obviously meant the semi-classical
model of measurement, when one can statistically infer the state of one
(quantum) part of a system immediately after observing the other (classical)
part, whatever the distance between them. In fact, there is no paradox of
``spooky action at distance'' in the classical case. The statistical
inference, playing the role of such immediate action, is simply based on the
Bayesian selection rule of a posterior state from the prior mixture of all
such states, corresponding to the possible results of the measurement. Bohr
always emphasized that one must treat the measuring instrument classically
(the measured spin, or another bit interacting with this spin, as a
classical bit), although the classical-quantum interaction should be
regarded as purely quantum. The latter follows from non-existence of
semi-classical Poisson bracket (i.e. classical-quantum potential
interaction). Schr\"{o}dinger clarified this point more precisely then Bohr,
and he followed in fact the mathematical pattern of von Neumann measurement
theory. EPR paradox is related to so called Bell inequality the
probabilistic roots of which was evidentiated in \cite{Acc81}.

\subsubsection{Releasing Schr\"{o}dinger's cat}

Motivated by EPR paper, in 1935 Schr\"{o}dinger published a three part essay 
\cite{Schr35} on `The Present Situation in Quantum Mechanics'. He turns to
EPR paradox and analyses completeness of the description by the wave
function for the entangled parts of the system. (The word \emph{entangled}
was introduced by Schr\"{o}dinger for the description of nonseparable
states.) He notes that if one has pure states $\psi \left( \sigma \right) $
and $\varphi \left( \tau \right) $ for each of two completely separated
bodies, one has maximal knowledge, $\chi \left( \sigma ,\tau \right) =\psi
\left( \sigma \right) \varphi \left( \tau \right) $, for two taken together.
But the converse is not true for the entangled bodies, described by a
non-separable wave function $\chi \left( \sigma ,\tau \right) \neq \psi
\left( \sigma \right) \varphi \left( \tau \right) $:

\begin{quote}
\textit{Maximal knowledge of a total system does not necessary imply maximal
knowledge of all its parts, not even when these are completely separated one
from another, and at the time can not influence one another at all.}
\end{quote}

To make absurdity of the EPR argument even more evident he constructed his
famous burlesque example in quite a sardonic style. A cat is shut up in a
steel chamber equipped with a camera, with an atomic mechanism in a pure
state $\rho_{0}=P_{\psi}$ which triggers the release of a phial of cyanide
if an atom disintegrates spontaneously (it is assumed that it might not
disintegrate in a course of an hour with probability $\mathrm{Tr}\left(
EP_{\psi}\right) =1/2$). If the cyanide is released, the cat dies, if not,
the cat lives. Because the entire system is regarded as quantum and closed,
after one hour, without looking into the camera, one can say that the entire
system is still in a pure state in which the living and the dead cat are
smeared out in equal parts.

Schr\"{o}dinger resolves this paradox by noting that the cat is a
macroscopic object, the states of which (alive or dead) could be
distinguished by a macroscopic observation as distinct from each other
whether observed or not. He calls this `the principle of state distinction'
for macroscopic objects, which is in fact the postulate that the directly
measurable system (consisting of the cat) must be classical:

\begin{quote}
\textit{It is typical in such a case that an uncertainty initially
restricted to an atomic domain has become transformed into a macroscopic
uncertainty which can be resolved through direct observation.}
\end{quote}

The dynamical problem of the transformation of the atomic, or ``coherent''
uncertainty, corresponding to a probability amplitude $\psi\left(
\sigma\right) $, into a macroscopic uncertainty, corresponding to a mixed
state $\rho$, is called quantum \emph{decoherence }problem$.$ In order to
make this idea clear, let us give the solution of the Schr\"{o}dinger's
elementary decoherence problem in the purely mathematical way. Instead of
the values $\pm1/2$ for the spin-variables $\sigma$ and $\tau$ we shall use
the values $\left\{ 0,1\right\} $ corresponding to the states of a classical
``bit'', the simplest nontrivial system in classical probability or
information theory.

Consider the atomic mechanism as a quantum ``bit'' with Hilbert space $%
\mathfrak{h}=\mathbb{C}^{2}$, the pure states of which are described by $%
\psi $-functions of the variable $\sigma \in \left\{ 0,1\right\} $ (if atom
is disintegrated, $\sigma =1$ , if not, $\sigma =0$) with scalar (complex)
values $\psi \left( \sigma \right) $ defining the probabilities $\left| \psi
\left( \sigma \right) \right| ^{2}$ of the quantum elementary propositions
corresponding to $\sigma =0,1$. The Schr\"{o}dinger's cat is a classical bit
with only two pure states $\tau \in \left\{ 0,1\right\} $ which can be
identified with the probability distributions $\delta _{0}\left( \tau
\right) $ when alive $\left( \tau =0\right) $ and $\delta _{1}\left( \tau
\right) $ when dead $\left( \tau =1\right) $. These and other (mixed) states
can also be described by the complex amplitudes $\varphi \left( \tau \right) 
$, however they are uniquely defined by the probabilities $\left| \varphi
\left( \tau \right) \right| ^{2}$ up to a phase function of $\tau $, the
phase multiplier of $\varphi \in \mathfrak{g}$, $\mathfrak{g}=\mathbb{C}^{2}$
commuting with all cat observables $c\left( \tau \right) $, not just up to a
phase constant as in the case of the atom (only constants commute with all
atomic observables $A\in \mathcal{L}\left( \mathfrak{h}\right) $). Initially
the cat is alive, so its amplitude (uniquely defined up to the phase factor
by the square root of probability distribution $\delta _{0}$ on $\left\{
0,1\right\} $) is $\delta \left( \tau \right) $ that is $1$ if $\tau =0$,
and $0$ if $\tau =1$.

The only meaningful classical-quantum reversible interaction affecting not
the atom but\ the cat after the hour, is described by unitary transformation 
\begin{equation*}
U\left[ \psi \otimes \varphi \right] \left( \sigma ,\tau \right) =\psi
\left( \sigma \right) \varphi \left( \left( \tau \bigtriangleup \sigma
\right) \right) ,
\end{equation*}
as both bits were quantum, where $\tau \bigtriangleup \sigma =\left| \tau
-\sigma \right| =\sigma \bigtriangleup \tau $ is the difference ($\func{mod}%
2 $) on $\left\{ 0,1\right\} $. Applied to the initial product-state $\psi
\otimes \delta $ it has the resulting probability amplitude 
\begin{equation*}
\chi \left( \sigma ,\tau \right) =\psi \left( \sigma \right) \delta \left(
\tau \bigtriangleup \sigma \right) =0\quad \mathrm{if}\quad \sigma \neq \tau
.
\end{equation*}
Despite the fact that the initial state was pure, $\chi _{0}=\psi \otimes
\delta $ corresponding to the Cartesian product $\left( \psi ,0\right) $ of
the initial pure states $\psi \in \mathfrak{h}$ and $\tau =0$, the
reversible unitary evolution $U$ induces in $\mathbb{H}=\mathfrak{h}\otimes 
\mathfrak{g}$ the mixed state $\chi \in \mathbb{H}$ for the
quantum-classical system ``atom+cat'' described by the wave function $\chi
\left( \sigma ,\tau \right) $.

Indeed, the observables of such a system are operator-functions $X$ of $\tau 
$ with values $X\left( \tau \right) $ in $\sigma $-matrices, represented as
block-diagonal $\left( \sigma ,\tau \right) $-matrices $\hat{X}=\left[
X\left( \tau \right) \delta _{\tau ^{\prime }}^{\tau }\right] $ of the
multiplication $X\left( \tau \right) \chi \left( \cdot ,\tau \right) $ at
each point $\tau \in \left\{ 0,1\right\} $. This means that the amplitude $%
\chi $ induces the same expectations 
\begin{equation*}
\left\langle \hat{X}\right\rangle =\sum_{\tau }\chi \left( \tau \right)
^{\dagger }X\left( \tau \right) \chi \left( \tau \right) =\sum_{\tau }%
\mathrm{Tr}X\left( \tau \right) \varrho \left( \tau \right) =\mathrm{Tr}\hat{%
X}\hat{\varrho}
\end{equation*}
as the block-diagonal density matrix $\hat{\varrho}=\left[ \varrho \left(
\tau \right) \delta _{\tau ^{\prime }}^{\tau }\right] $ of the
multiplication by 
\begin{equation*}
\varrho \left( \tau \right) =E\left( \tau \right) P_{\psi }E\left( \tau
\right) =\pi \left( \tau \right) P_{E\left( \tau \right) \psi }
\end{equation*}
where $\pi \left( \tau \right) =\left| \psi \left( \tau \right) \right| ^{2}$%
, $E\left( \tau \right) =P_{\delta _{\tau }}$ is the projection operator 
\begin{equation*}
\left[ E\left( \tau \right) \psi \right] \left( \sigma \right) =\delta
\left( \tau \bigtriangleup \sigma \right) \psi \left( \sigma \right) =\psi
\left( \tau \right) \delta _{\tau }\left( \sigma \right) ,
\end{equation*}
and $P_{E\left( \tau \right) \psi }=P_{\delta _{t}}$ is also the projector
onto $\delta _{\tau }\left( \cdot \right) =\delta \left( \cdot
\bigtriangleup \tau \right) $. The $4\times 4$-matrix $\hat{\varrho}$ is a
mixture of two orthogonal projectors $P_{\delta _{\tau }}\otimes P_{\delta
_{\tau }}$, $\tau =0,1$: 
\begin{equation*}
\hat{\varrho}=\left[ P_{\delta _{\tau }}\delta _{\tau ^{\prime }}^{\tau }\pi
\left( \tau \right) \right] =\sum_{\tau =0}^{1}\pi \left( \tau \right)
P_{\delta _{\tau }}\otimes P_{\delta _{\tau }}.
\end{equation*}

\subsubsection{Von Neumann's projection postulate}

Inspired by Bohr's complementarity principle, von Neumann proposed even
earlier the idea that every quantum measuring process involves an
unanalysable element. He postulated \cite{Neum32} that, in addition to the
continuous causal propagation of the wave function generated by the Schr\"{o}%
dinger equation, during a measurement, due to an action of the observer on
the object, the function undergoes a discontinuous, irreversible
instantaneous change. Thus, just prior to a measurement of the event $F$,
disintegration of the atom, say, the quantum pure state $P_{\psi }$ changes
to the mixed one 
\begin{equation*}
\rho =\lambda P_{E\psi }+\mu P_{F\psi }=E\rho E+F\rho F,
\end{equation*}
where $E=I-F$ is the orthocomplement event, and $\lambda =\mathrm{Tr}E\rho $%
, $\mu =\mathrm{Tr}F\rho $ are the probabilities of $E$ and $F$. Such change
is projective as shown in the second part of this equation, and it is called
the von Neumann projection postulate.

This linear irreversible decoherence process should be completed by the
nonlinear, acausal random jump to one of the pure states 
\begin{equation*}
\rho \mapsto P_{E\psi }\text{, or }\rho \mapsto P_{F\psi }
\end{equation*}
depending on whether the tested event $F$ is false (the cat is alive, $\psi
_{0}=\lambda ^{-1/2}E\psi $), or true (the cat is dead, $\psi _{1}=\mu
^{-1/2}F\psi $). This final step is the posterior prediction, called \emph{%
filtering} of the decoherent mixture of $\psi _{0}$ and $\psi _{1}$ by
selection of only one result of the measurement, and is an unavoidable
element in every measurement process relating the state of the pointer of
the measurement (in this case the cat) to the state of the whole system.
This assures that the same result would be obtained in case of immediate
subsequent measurement of the same event $F$. The resulting change of the
initial wave-function $\psi $ is described up to normalization by one of the
projections 
\begin{equation*}
\psi \mapsto E\psi ,\quad \psi \mapsto F\psi
\end{equation*}
and is sometimes called the L\"{u}ders projection postulate \cite{Lud51}.

Although unobjectionable from the purely logical point of view the von
Neumann theory of measurement soon became the target of severe criticisms.
Firstly it seams radically subjective, postulating the spooky action at
distance (the filtering) in a purely quantum system instead of deriving it.
Secondly the analysis is applicable to only the idealized situation of
discrete instantaneous measurements.

However as we already mentioned when discussing the EPR paradox, the process
of filtering is free from conceptual difficulty if it is understood as the
statistical inference about a mixed state in an extended stochastic
representation of the quantum system as a part of a semiclassical one, based
upon the results of observation in its classical part. In order to
demonstrate this, we can return to the dynamical model of Schr\"{o}dinger's
cat, identifying the quantum system in question with the Schr\"{o}dinger's
atom. The event $E$ (the atom exists) corresponds then to $\tau =0$ (the cat
is alive), $E=E\left( 0\right) $, and the complementary event is $F=E\left(
1\right) $. This model explains that the origin of the von Neumann
irreversible decoherence $P_{\psi }\mapsto \rho $ of the atomic state is in
the ignorance of the result of the measurement described by the partial
tracing over the cat's Hilbert space $\mathfrak{g}=\mathbb{C}^{2}$: 
\begin{equation*}
\rho =\mathrm{Tr}_{\mathfrak{g}}\hat{\varrho}=\sum_{\tau =0}^{1}\pi \left(
\tau \right) P_{\delta _{\tau }}=\varrho \left( 0\right) +\varrho \left(
1\right) ,
\end{equation*}
where $\varrho \left( \tau \right) =\left| \psi \left( \tau \right) \right|
^{2}P_{\delta _{\tau }}$. It has entropy $S\left( \rho \right) =\mathrm{Tr}%
\rho \log \rho ^{-1}$ of the compound state $\hat{\varrho}$ of the combined
semi-classical system prepared for the indirect measurement of the
disintegration of atom by means of cat's death: 
\begin{equation*}
S\left( \rho \right) =-\sum_{\tau =0}^{1}\left| \psi \left( \tau \right)
\right| ^{2}\log \left| \psi \left( \tau \right) \right| ^{2}=S\left( \hat{%
\varrho}\right)
\end{equation*}
It is the initial coherent uncertainty in the pure quantum state of the atom
described by the wave-function $\psi $ which is equal to one bit if
initially $\left| \psi \left( 0\right) \right| ^{2}=1/2=\left| \psi \left(
1\right) \right| ^{2}$.

This dynamical model of the measurement which is due to von Neumann, also
interprets the filtering $\rho\mapsto\rho_{\tau}$ simply as the conditioning 
\begin{equation*}
\rho_{\tau}=\varrho\left( \tau\right) /\pi\left( \tau\right)
=P_{\delta_{\tau}}
\end{equation*}
of the joint classical-quantum state $\varrho\left( \cdot\right) $ by the
Bayes formula which is applicable due to the commutativity of actually
measured observable (the life of cat) with any other observable of the
combined semi-classical system.

Thus the atomic decoherence is derived from the unitary interaction of the
quantum atom with the classical cat. The spooky action at distance,
affecting the atomic state by measuring $\tau$, is simply the result of the
statistical inference (prediction after the measurement) of the atomic
posterior state $\rho_{\tau}=P_{\delta_{\tau}}$: the atom disintegrated if
and only if the cat is dead. A formal derivation of the von-Neumann-L\"{u}%
ders projection postulate and the decoherence in the general case by
explicit construction of unitary transformation in the extended
semi-classical system is given in \cite{Be94, StBe96}.

\section{Decoherence, Measurement and Filtering}

\medskip\medskip

\textit{In mathematics you don't understand things. You just get used to them%
} - John von Neumann.

In this Chapter I present the author's views on the solution to quantum
measurement problem which might not coincide with the present scientific
consensus that this problem is unsolvable, or at least unsolved. It will be
shown that there exists such solution along the line suggested by the great
founders of quantum theory Schr\"{o}dinger, Heisenberg and Bohr. In fact it
was envisaged by von Neumann in his Mathematical Foundation of Quantum
Theory \cite{Neum32}, and by more recent quantum philosopher J Bell in \cite%
{Bell87}. However while the dynamical consideration of the measurement
process in \cite{Neum32} is absent at all, the continuous in time model of
measurement suggested by Bell is simply wrong (the dynamical equation (5) he
suggested in \cite{Bell87}, p.176, doesn't preserve the positivity of
transition probabilities for the stochastic process).

The differential analysis of the appropriate models is based on It\^{o}
stochastic calculus and its quantum generalization. The discovery of quantum
thermal noise and its white-noise approximations lead to a profound
revolution not only in modern physics but also in contemporary mathematics
comparable with the discovery of differential calculus by Newton (for a
feature exposition of this, accessible for physicists, see \cite{Gar91}, the
complete theory, which was mainly developed in the 80's \cite{Be80, HuPa84,
GaCo85, Be88a}, is sketched in the Appendix 2) .

\subsection{Beables and Nondemolition Principle}

Schr\"{o}dinger like Einstein was deeply concerned with the loss of reality
and causality in the positivistic treatment of quantum measuring process by
Heisenberg and Born. Schr\"{o}dinger's remained unhappy with Bohr's reply to
the EPR paradox, Schr\"{o}dinger's own analysis was:

\begin{quote}
\textit{It is pretty clear, if reality does not determine the measured
value, at least the measurable value determines reality.}
\end{quote}

Our approach resolves the famous paradoxes of quantum measurement theory in
a constructive way by giving exact nontrivial models for the statistical
analysis of quantum observation processes determining the reality underlying
these paradoxes. Conceptually it is based upon a new idea of quantum
causality called the Nondemolition Principle \cite{Be94} which divides the
world into the classical past, forming the consistent histories, and the
quantum future, the state of which is predictable for each such history.

\subsubsection{Compatibility and time arrow}

Von Neumann's projection postulate and its dynamical realization can be
generalized to include cases with continuous spectrum of values. In fact
there many such developments, we will only mention here the most general
operational approach to quantum measurements of Ludwig \cite{Lud68}, and its
mathematical implementation by Davies and Lewis \cite{DaLe70} in the
``instrumental'' form. The stochastic realization of the corresponding
completely positive reduction map $\rho \mapsto \varrho \left( \cdot \right) 
$, resolving the corresponding instantaneous quantum measurement problem,
can be found in \cite{Oz84, Be94}. Because of the crucial importance of
these realizations for developing understanding of the mathematical
structure and interpretation of modern quantum theory, we need to analyze
the mathematical consequences which can be drawn from such schemes.

The generalized reduction of the wave function $\psi \left( x\right) $,
corresponding to a complete measurement with discrete or continuous data $y$%
, is described by a function $V\left( y\right) $ whose values are linear
operators $\mathfrak{h}\ni \psi \mapsto V\left( y\right) \psi $ which are,
in general, not isometric on the given Hilbert space $\mathfrak{h}$, $%
V\left( y\right) ^{\ast }V\left( y\right) \neq I$, but have the following
normalization condition. The resulting wave-function 
\begin{equation*}
\chi \left( x,y\right) =\left[ V\left( y\right) \psi \right] \left( x\right)
\end{equation*}
is normalized with respect to a given measure $\mu $ on $y$ in the sense 
\begin{equation*}
\iint \left| \chi \left( x,y\right) \right| ^{2}\mathrm{d}\mu \mathrm{d}%
\lambda =\int \left| \psi \left( x\right) \right| ^{2}\mathrm{d}\lambda
\end{equation*}
for any probability amplitude $\psi $ (normalized with respect to a measure $%
\lambda $). This can be written as $V^{\dagger }V=I$ in terms of the
integral 
\begin{equation*}
\int_{y}V\left( y\right) ^{\ast }V\left( y\right) \mathrm{d}\mu =I,\quad 
\mathrm{or}\quad \sum_{y}V\left( y\right) ^{\ast }V\left( y\right) =I.
\end{equation*}
with respect to the base measure $\mu $ which is taken in the discrete case,
such as the case of two-point variables $y=\tau $ (EPR paradox, or Schr\"{o}%
dinger cat with the projection-valued $V\left( \tau \right) =E\left( \tau
\right) $), to be the counting measure. As in that simple example the
realization of such $V$ can be always constructed \cite{Be94} in terms of a
unitary transformation $U$ on an extended Hilbert space $\mathfrak{h}\otimes 
\mathfrak{g}$ and a normalized wave function $\varphi _{0}\in \mathfrak{g}$
such that 
\begin{equation*}
U\left[ \psi \otimes \varphi _{0}\right] \left( x,y\right) =\chi \left(
x,y\right)
\end{equation*}
for any $\psi $. The additional system described by ``the pointer coordinate 
$y$ of the measurement apparatus'' can be regarded as classical (like the
cat) as the actual observables in question are the measurable functions $%
g\left( y\right) $ represented by commuting operators $\hat{g}$ of
multiplication by these functions. They are appropriate candidates for
Bell's ''beables'', \cite{Bell87}, p.174, as such commuting observables,
extended to the quantum part as $I\otimes \hat{g}$, are compatible with any
possible (future) event, represented by an orthoprojector $F\otimes I$. The
probabilities (or, it is better to say, the propensities) of all such events
are the same in all states whether an observable $\hat{g}$ was measured but
the result is not read, or it was not measured at all. In this sense the
measurements of $\hat{g}$ are called \textit{nondemolition} with respect to
the future observables $F$, they do not demolish the picture of the
possibilities, or propensities of $F$. But they are not necessary compatible
with the initial operators $F\otimes I$ of the quantum system under the
question in the present representation $U\left( F\otimes I\right) U^{\ast }$
corresponding to the actual states $\chi =U\left( \psi \otimes \varphi
_{0}\right) $.

Indeed, the Heisenberg operators 
\begin{equation*}
G=U^{\ast}\left( I\otimes\hat{g}\right) U
\end{equation*}
of the nondemolition observables in general do not commute with the past
operators $F\otimes I$ on the initial states $\chi_{0}=\psi\otimes%
\varphi_{0} $. One can see this from the example of the Schr\"{o}dinger cat.
The ``cat observables'' in Heisenberg picture are represented by commuting
operators $G=\left[ g\left( \sigma+\tau\right)
\delta_{\sigma^{\prime}}^{\sigma }\delta_{\tau^{\prime}}^{\tau}\right] $ of
multiplication by $g\left( \sigma+\tau\right) $, where the sum $%
\sigma+\tau=\left| \sigma-\tau\right| $ is modulo 2. They do not commute
with $F\otimes I$ unless $F$ is also a diagonal operator $\hat{f}$ of
multiplication by a function $f\left( \sigma\right) $ in which case 
\begin{equation*}
\left[ F,G\right] \chi_{0}\left( \sigma,\tau\right) =\left[ f\left(
\sigma\right) ,g\left( \sigma+\tau\right) \right] \chi_{0}\left(
\sigma,\tau\right) =0,\quad\forall\chi_{0}\text{.}
\end{equation*}
However the restriction of the possibilities in a quantum system to only the
diagonal operators $F=\hat{f}$ which would eliminate the time arrow in the
nondemolition condition amounts to the redundancy of the quantum
consideration as all such (possible and actual) observables can be
simultaneously represented as the functions of $\left( \sigma,\tau\right) $
as in the classical case.

\subsubsection{Transition from possible to actual}

The analysis above shows that as soon as dynamics is taken into
consideration even in the form of just a single unitary transformation, the
measurement process needs to specify the arrow of time, what is the
predictable future and what is the reduced past, what is possible and what
is actual with respect to this measurement. As soon as a measured observable 
$Y$ is specified, i.e. is taken as a \textit{be}able, all other operators
which do not commute with $Y$ become entirely redundant and are not among
possible future \textit{be}ables. The algebra $\mathcal{A}$ of all such
potential future observables (not the state which stays invariant in the Schr%
\"{o}dinger picture unless the selection due to an inference has taken
place!) reduces to the subalgebra commuting with $Y$, and \emph{this
reduction doesn't change the reality} (the wave function remains the same
and induces the same, now mixed, state on the smaller, reduced algebra!).
Possible observables in an individual system are only those which are
compatible with the actual observable/beables. This is another formulation
of Bohr's complementarity which specifies \emph{mathematically} which
natural processes have the special status of `measurements', and which was
unknown to Bell (compare with \textquotedblleft There is nothing in the
mathematics to tell what is `system' and what is `apparatus',
...\textquotedblright , in \cite{Bell87}, p.174). More specifically this can
be rephrased in the form of a dynamical postulate of quantum causality
called the Nondemolition Principle \cite{Be94} which we first formulate for
a single instant of time $t$ in quite an obvious form:

\begin{quote}
In the appropriate representation of a quantum system by an algebra $%
\mathcal{A}$ of (necessarily not all) operators on a Hilbert space of the
system plus measurement apparatus, causal, or nondemolition observables are
represented only by those operators $Y$, which are compatible with $\mathcal{%
A}$: 
\begin{equation*}
\left[ X,Y\right] :=XY-YX=0,\quad \forall X\in \mathcal{A}
\end{equation*}
(this is usually written as $Y\in \mathcal{A}^{\prime }$, where $\mathcal{A}%
^{\prime }$, called the commutant of $\mathcal{A}$, in this formulation is
not necessarily contained in $\mathcal{A}$). Each measurement process of the
history for a quantum system $\mathcal{A}$ can be represented as
nondemolition by the causal observables in the appropriate representation of 
$\mathcal{A}$.
\end{quote}

Note that the space of representation plays here the crucial role: the
reduced operators 
\begin{equation*}
X_{0}=\left( I\otimes \varphi _{0}\right) ^{\ast }X\left( I\otimes \varphi
_{0}\right) ,\;Y_{0}=\left( I\otimes \varphi _{0}\right) ^{\ast }Y\left(
I\otimes \varphi _{0}\right)
\end{equation*}
for commuting $X$ and $Y$ might not commute on the smaller space $\mathfrak{h%
}_{0}\subset \mathfrak{h}\otimes \mathfrak{g}$ of the initial states $\psi
\otimes \varphi _{0}$ with a fixed $\varphi _{0}\in \mathfrak{g}$. Even if
the nondemolition observables $Y$ is faithfully represented by $Y_{0}$ on
initial space $\mathfrak{h}_{0}$, as it is in the case $Y=G$ of the Schr\"{o}%
dinger's cat with $\varphi _{0}\left( \tau \right) =\delta \left( \tau
\right) $, where $Y_{0}$ is the multiplication operator $G_{0}=\hat{g}$ for $%
\psi $: 
\begin{equation*}
G\left( \psi \otimes \delta \right) \left( \sigma ,\tau \right) =g\left(
\sigma +\tau \right) \psi \left( \sigma \right) \delta \left( \tau \right)
=g\left( \sigma \right) \psi \left( \sigma \right) \delta \left( \tau
\right) =\left( G_{0}\psi \otimes \delta \right) \left( \sigma ,\tau \right)
,
\end{equation*}
there is usually no room in $\mathfrak{h}_{0}$ to represent all Heisenberg
operators $X\in \mathcal{A}$ commuting with $Y$ on $\mathfrak{h}\otimes 
\mathfrak{g}$. The induced operators $Y_{0}$ do not commute with all
operators $F$ of the system initially represented on $\mathfrak{h}_{0}$, and
this is why the measurement of $Y_{0}$ is thought to cause demolition on $%
\mathfrak{h}_{0}$. However in all such cases the future operators $X$
reduced to $X_{0}$ on $\mathfrak{h}_{0}$,\ commute with $Y_{0}$ as they are
decomposable with respect to $Y_{0}$ (however the reduction $X\mapsto X_{0}$
is not the Heisenberg one-to-one but instead an irreversible dynamical map).
This can be seen explicitly for the atom described by the Heisenberg
operators $X=U^{\ast }\left( F\otimes I\right) U$ in the interaction
representation with the cat: 
\begin{equation*}
X_{0}=\sum_{\tau }E\left( \tau \right) FE\left( \tau \right) ,\quad
Y_{0}=\sum g\left( \tau \right) E\left( \tau \right) .
\end{equation*}

The nondemolition principle can be considered not only as a restriction on
the possible observations for a given dynamics but also as a condition for
the causal dynamics to be compatible with the given observations (beables).
As was proved in \cite{Be92a}, the causality condition is necessary and
sufficient for the existence of a conditional expectation for any state on
the total algebra $\mathcal{A}\vee \mathcal{B}$ with respect to a
commutative subalgebra $\mathcal{B}$ of nondemolition observables $Y$. Thus
the nondemolition causality condition amounts exactly to the existence of
the conditional states, i.e. to the predictability of the states on the
algebra $\mathcal{A}$ upon the measurement results of the observables in $%
\mathcal{B}$. Then the transition from a prior $\rho $ to a posterior state $%
\rho _{y}=P_{V\left( y\right) \psi }$ is simply the result of gaining
knowledge of $y$ defining the actual state in the decoherent mixture 
\begin{equation*}
\rho =\int V\left( y\right) P_{\psi }V\left( y\right) ^{\ast }\mathrm{d}\mu
=\int P_{V\left( y\right) \psi }f\left( y\right) \mathrm{d}\mu
\end{equation*}
of all possible states, where $f\left( y\right) =\left\| V\left( y\right)
\psi \right\| ^{2}$ is the probability density of $y$ defining the output
measure $\mathrm{d}\nu =f\mathrm{d}\mu $. As Heisenberg always emphasized,
``quantum jump'' is contained in the transitions from possible to actual.

If an algebra $\mathcal{B}$ of beables is specified at a time $t$, there
must be a causal representation $\mathcal{B}_{t}$ of $\mathcal{B}$ with
respect to the present $\mathcal{A}_{t}$ and all future possible
representations $\mathcal{A}_{s}$, $s>t$ of the quantum system on the same
Hilbert space (they might not coincide with $\mathcal{A}_{t}$ if the system
is open \cite{AFL82}). The past representations $\mathcal{A}_{r}$, $r<t$
which are incompatible with a $G\in\mathcal{B}_{t}$ are meaningless as
noncausal for the observation at the time $t$, they should be replaced by
the causal histories $\mathcal{B}_{r}$, $r<t$ of the beables which must be
consistent in the sense of compatibility of all $\mathcal{B}_{t}$. Thus the
dynamical formulation of the nondemolition principle of quantum causality
and the consistency of histories reads as 
\begin{equation*}
\mathcal{A}_{s}\subseteq\mathcal{B}_{r}^{\prime},\quad\mathcal{B}%
_{s}\subseteq\mathcal{B}_{r}^{\prime},\quad\forall r\leq s.
\end{equation*}

These are the only possible conditions when the posterior states always
exist as results of inference (filtering and prediction) of future quantum
states upon the measurement results of the classical (i.e. commutative) past
of a process of observation. The act of measurement transforms quantum
propensities into classical realities. As Lawrence Bragg, another Nobel
prize winner, once said, everything in the future is a wave, everything in
the past is a particle.

\subsubsection{The true Heisenberg principle}

The time continuous solution of the quantum measurement problem was
motivated by analogy with the classical stochastic filtering problem which
obtains the prediction of future for an unobservable dynamical process $%
x\left( t\right) $ by time-continuous measuring of another, observable
process $y\left( t\right) $. Such a problem was first considered by Wiener
and Kolmogorov who found its solution in the form of \ causal spectral
filter but only for the stationary Gaussian case. The differential solution
in the form of a stochastic filtering equation was then obtained by
Stratonovich \cite{Str66} in 1958 for an arbitrary Markovian pair $\left(
x,y\right) $. This was really a break through in the statistics of
stochastic processes which soon found many applications, in particular for
solving the problems of stochastic control under incomplete information (it
is possible that this was one of the reasons why the Russians were so
successful in launching the rockets to the Moon and other planets of the
Solar system in 60s).

If $X\left( t\right) $ is the unobservable process, a Heisenberg coordinate
process of a quantum particle, say, and $Y\left( t\right) $ is an observable
quantum process, describing the trajectories $y\left( t\right) $ of the
particle in a cloud chamber, say, why don't we find a filtering equation for
the a posterior expectation $q\left( t\right) $ of $X\left( t\right) $ or
any other \ function of $X\left( t\right) $ in the same way as we do it in
the classical case if we know a history, i.e. a particular trajectory $%
y\left( r\right) $ up to the time $t$? This problem was first considered and
solved for the case of quantum Markovian Gaussian pair $\left( X,Y\right) $
corresponding to a quantum open linear system with linear output channel, in
particular for a quantum oscillator matched to a quantum transmission line 
\cite{Be80, Be85}. By studying this example, the nondemolition condition 
\begin{equation*}
\left[ X\left( s\right) ,Y\left( r\right) \right] =0,\quad \text{ }\left[
Y\left( s\right) ,Y\left( r\right) \right] =0\quad \forall r\leq s
\end{equation*}
was first found, and this allowed the solution in the form of the causal
equation for $q\left( t\right) =\left\langle X\left( t\right) \right\rangle
_{y}$.

Let us describe this exact dynamical model of the causal nondemolition
measurement first in terms of quantum white noise analysis for a
one-dimensional quantum nonrelativistic particle of mass $m$ which is
conservative if not observed, in a potential field $\phi $. But we shall
assume that the particle is under indirect observation by measuring of its
Heisenberg position operator $X\left( t\right) $ with an additive random
error $e\left( t\right) :$%
\begin{equation*}
Y\left( t\right) =X\left( t\right) +e\left( t\right) .
\end{equation*}
We take the simplest statistical model for the error process $e\left(
t\right) $, the white noise model (the worst, completely chaotic error),
assuming that it is a classical (i.e. commutative) Gaussian white noise
given by the first momenta 
\begin{equation*}
\left\langle e\left( t\right) \right\rangle =0,\quad \left\langle e\left(
s\right) e\left( r\right) \right\rangle =\sigma ^{2}\delta \left( s-r\right)
.
\end{equation*}
The measurement process $Y\left( t\right) $ should be commutative,
satisfying the causal nondemolition condition with respect to the
noncommutative process $X\left( t\right) $ (and any other Heisenberg
operator-process of the particle), this can be achieved by perturbing the
particle Newton-Erenfest equation: 
\begin{equation*}
m\frac{\mathrm{d}^{2}}{\mathrm{d}t^{2}}X\left( t\right) +\nabla \phi \left(
X\left( t\right) \right) =f\left( t\right) .
\end{equation*}
Here $f\left( t\right) $ is the Langevin force perturbing the dynamics due
to the measurement, which is assumed to be another classical (commutative)
white noise. 
\begin{equation*}
\left\langle f\left( t\right) \right\rangle =0,\quad \left\langle f\left(
s\right) f\left( r\right) \right\rangle =\tau ^{2}\delta \left( s-r\right) .
\end{equation*}
In classical measurement and filtering theory the white noises $e\left(
t\right) ,f\left( t\right) $ are usually considered independent, and the
intensities $\sigma ^{2}$ and $\tau ^{2}$ can be arbitrary, even zeros,
corresponding to the ideal case of the direct unperturbing observation of
the particle trajectory $X\left( t\right) $. However in quantum theory
corresponding to the standard commutation relations 
\begin{equation*}
X\left( 0\right) =\hat{x},\quad \frac{\mathrm{d}}{\mathrm{d}t}X\left(
0\right) =\frac{1}{m}\hat{p},\quad \left[ \hat{x},\hat{p}\right] =i\hbar 
\hat{1}
\end{equation*}
the particle trajectories do not exist, and it was always understood that
the measurement error $e\left( t\right) $ and perturbation force $f\left(
t\right) $ should satisfy a sort of uncertainty relation. This ``true
Heisenberg principle'' had never been mathematically formulated and proved
before the discovery \cite{Be80} of quantum causality and nondemolition
condition in the above form of commutativity of $X\left( s\right) $ and $%
Y\left( r\right) $ for $r\leq s$. As we showed first in the linear case \cite%
{Be80, Be85}, and later even in the most general case \cite{Be92b}, these
conditions are fulfilled if and only if $e\left( t\right) $ and $f\left(
t\right) $ satisfy the canonical commutation relations 
\begin{equation*}
\left[ e\left( r\right) ,e\left( s\right) \right] =0,\;\left[ e\left(
r\right) ,f\left( s\right) \right] =\frac{\hbar }{i}\delta \left( r-s\right)
,\;\left[ f\left( r\right) ,f\left( s\right) \right] =0.
\end{equation*}
This proves that the pair $\left( e,f\right) $ must satisfy the uncertainty
relation $\sigma \tau \geq \hbar /2$, i.e. 
\begin{equation*}
\Delta e_{t}\Delta f_{t}\geq \hbar t/2\text{,}
\end{equation*}
in terms of the standard deviations of the integrated processes 
\begin{equation*}
e_{t}=\int_{0}^{t}e\left( r\right) \mathrm{d}r,\quad
f_{t}=\int_{0}^{t}f\left( s\right) \mathrm{d}s.
\end{equation*}
This inequality constitutes the precise formulation of the true Heisenberg
principle for the square roots $\sigma $ and $\tau $ of the intensities of
error $e$ and perturbation $f$: they are inversely proportional with the
same coefficient of proportionality, $\hbar /2$, as for the pair $\left( 
\hat{x},\hat{p}\right) $. The canonical pair $\left( e,f\right) $ called
quantum white noise cannot be considered classically despite of the
possibility of the classical realizations of each process $e$ and $f$
separately due to the self-commutativity of the families $e$ and $f$. Thus,
a generalized matrix mechanics for the treatment of quantum open systems
under continuous nondemolition observation and the true Heisenberg principle
was invented exactly 20 years ago in \cite{Be80}. The nondemolition
commutativity of $Y\left( t\right) $ with respect to the Heisenberg
operators of the open quantum system was later rediscovered for the output
of quantum stochastic fields in \cite{GaCo85}.

\subsection{Consistent Histories and Filtering}

Schr\"{o}dinger believed that all quantum problems including the
interpretation of measurement should be formulated in continuous time in the
form of differential equations. He thought that the measurement problem
would have been resolved if quantum mechanics had been made consistent with
relativity theory and the time had been treated appropriately. However
Einstein and Heisenberg did not believe this, each for his own reasons.
While Einstein thought that the probabilistic interpretation of quantum
mechanics was wrong, Heisenberg simply stated:-

\begin{quote}
\textit{Quantum mechanics itself, whatever its interpretation, does not
account for the transition from `possible to the actual'}
\end{quote}

Perhaps the closest to the truth was Bohr when he said that it `must be
possible so to describe the extraphysical process of the subjective
perception as\ if it were in reality in the physical world', extending the
reality beyond the closed quantum mechanical form by including a subjective
observer into a semiclassical world. He regarded the measurement apparatus,
or meter, as a semiclassical object which interacts with the world in a
quantum mechanical way but has only commuting observables - pointers. Thus
Bohr accepted that \textit{not all the world is quantum mechanical, there is
a classical part of the physical world, and we belong partly to this
classical world.}

In realizing this program I will follow the line suggested by John Bell \cite%
{Bell87} along which the ``development towards greater physical precision
would be to have the `jump' in the equations and not just the talk -- so
that it would come about as a dynamical process in dynamically defined
conditions.''

\subsubsection{Stochastic decoherence equation}

The generalized wave mechanics which enables us to treat the quantum
processes of time continuous observation, or in other words, quantum
mechanics with trajectories $\omega =\left( y_{t}\right) $, was discovered
only quite recently, in \cite{Be88, Be89a, Be89b}. The basic idea of the
theory is to replace the deterministic unitary Schr\"{o}dinger propagation $%
\psi \mapsto \psi \left( t\right) $ by a linear causal stochastic one $\psi
\mapsto \chi \left( t,\omega \right) $ which is not necessarily unitary for
each history $\omega $, but unitary in the mean square sense with respect to
a standard probability measure $\mu \left( \mathrm{d}\omega \right) $ for $%
\mathrm{d}\omega $. Due to this the positive measures 
\begin{equation*}
\mathrm{P}\left( t,\mathrm{d}\omega \right) =\left\| \chi \left( t,\omega
\right) \right\| ^{2}\mu \left( \mathrm{d}\omega \right) ,\quad \tilde{\mu}%
\left( \mathrm{d}\omega \right) =\lim_{t\rightarrow \infty }\mathrm{P}\left(
t,\mathrm{d}\omega \right)
\end{equation*}
are normalized (if $\left\| \psi \right\| =1$) for each $t$, and are
interpreted as the probability measure for the histories $\omega _{t}=\left(
y_{r}\right) _{r<t}$.of the output stochastic process $y_{t}$ with respect
to the measure $\tilde{\mu}$. In the same way as the abstract Schr\"{o}%
dinger equation can be derived from only unitarity of propagation, the
abstract decoherence wave equation can be derived from the mean square
unitarity in the form of a linear stochastic differential equation. The
reason that Bohr and Schr\"{o}dinger didn't derive such an equation despite
their firm belief that the measurement process can be described `as\ if it
were in reality in the physical world' is that the appropriate (stochastic)
differential calculus had not been yet developed early in that century. As
Newton had to invent the differential calculus in order to formulate the
equations of classical dynamics, we had to develop the quantum stochastic
calculus for nondemolition processes \cite{Be88, Be92a} presented in the
Appendix in order to derive the generalized wave equation for quantum
dynamics with continual observation.

For the notational simplicity we shall consider here the one dimensional
case, $d=1$, the multi-dimensional case is discussed in the Appendix 2 and
can be found elsewhere (e.g. in \cite{Be88, Be92a}). The abstract stochastic
wave equation can be written in this case as 
\begin{equation*}
\mathrm{d}\chi\left( t\right) +K\chi\left( t\right) \mathrm{d}t=L\chi\left(
t\right) \mathrm{d}y_{t},\quad\chi\left( 0\right) =\psi.
\end{equation*}
Here $y_{t}\left( \omega\right) $ is assumed to be a martingale (e.g. the
independent increment process with zero expectation, see the Appendix)
representing a measurement noise with respect to the input probability
measure $\mu\left( \mathrm{d}\omega\right) =\mathrm{P}\left( 0,\mathrm{d}%
\omega\right) $ (but not with respect to the output probability measure $%
\tilde{\mu}=\mathrm{P}\left( \infty,\mathrm{d}\omega\right) $ for which $%
y_{t}\left( \omega\right) $ is an output process with dependent increments).
If the stochastic process $\chi\left( t,\omega\right) $ is normalized in the
mean square sense for each $t$, it represents a probability amplitude $%
\chi\left( t\right) $ in an extended Hilbert space describing the process of
continual decoherence of the initial pure state $\rho\left( 0\right)
=P_{\psi}$ into the mixture 
\begin{equation*}
\rho\left( t\right) =\int P_{\psi_{\omega}\left( t\right) }\mathrm{P}\left(
t,\mathrm{d}\omega\right) =\mathrm{M}\left[ \chi\left( t\right) \chi\left(
t\right) ^{\dagger}\right]
\end{equation*}
of the posterior states corresponding to $\psi_{\omega}\left( t\right)
=\chi\left( t,\omega\right) /\left\| \chi\left( t,\omega\right) \right\| $,
where $\mathrm{M}$ denotes mean with respect to the measure $\mu$. Assuming
that the conditional expectation $\left\langle \mathrm{d}y_{t}\mathrm{d}%
y_{t}\right\rangle _{t}$ in 
\begin{align*}
\left\langle \mathrm{d}\left( \chi^{\dagger}\chi\right) \right\rangle _{t} &
=\left\langle \mathrm{d}\chi^{\dagger}\mathrm{d}\chi+\chi^{\dagger }\mathrm{d%
}\chi+\mathrm{d}\chi^{\dagger}\chi\right\rangle _{t} \\
& =\chi^{\dagger}\left( L^{\ast}\left\langle \mathrm{d}y_{t}\mathrm{d}%
y_{t}\right\rangle _{t}L-\left( K+K^{\ast}\right) \mathrm{d}t\right) \chi
\end{align*}
is $\mathrm{d}t$ (e.g. $\left( \mathrm{d}y_{t}\right) ^{2}=\mathrm{d}%
t+\varepsilon\mathrm{d}y_{t}$), the mean square normalization in its
differential form $\left\langle \mathrm{d}\left( \chi^{\dagger}\chi\right)
\right\rangle _{t}=0$ can be expressed \cite{Be89a, Be90b} as $K+K^{\ast
}=L^{\ast}L$, \ i.e. 
\begin{equation*}
K=\frac{1}{2}L^{\ast}L+\frac{i}{\hbar}H,
\end{equation*}
where $H=H^{\ast}$ is the Schr\"{o}dinger Hamiltonian such that this is the
Schr\"{o}dinger equation if $L=0$. One can also derive the corresponding
Master equation 
\begin{equation*}
\frac{\mathrm{d}}{\mathrm{d}t}\rho\left( t\right) +K\rho\left( t\right)
+\rho\left( t\right) K^{\ast}=L\rho\left( t\right) L^{\ast}
\end{equation*}
for mixing decoherence of the initially pure state $\rho\left( 0\right)
=\psi\psi^{\dagger}$, as well as a stochastic nonlinear wave equation for
the dynamical prediction of the posterior state vector $\psi_{\omega}\left(
t\right) $, the normalization of $\chi\left( t,\omega\right) $ at each $%
\omega$.

\subsubsection{Quantum jumps and diffusions}

Actually, there are two basic standard forms \cite{Be89b, Be90a} of such
stochastic wave equations, corresponding to two basic types of stochastic
integrators with independent increments: the Brownian standard type, $%
\varepsilon =0$, $y_{t}\simeq w_{t}$, and the Poisson standard type $%
\varepsilon =1$, $y_{t}\simeq n_{t}-t$ with respect to the basic measure $%
\mu $, see the Appendix. To get these we shall assume that $y_{t}$ is
standard with respect to the input measure $\mu $, given by the
multiplication table 
\begin{equation*}
\left( \mathrm{d}y\right) ^{2}=\mathrm{d}t+\nu ^{-1/2}\mathrm{d}y,\quad 
\mathrm{d}y\mathrm{d}t=0=\mathrm{d}t\mathrm{d}y,
\end{equation*}
where $\nu >0$ is the intensity of the Poisson process $n_{t}=\nu
^{1/2}y_{t}+\nu t$, and 
\begin{equation*}
L=\nu ^{1/2}(C-I),\quad H=E+i\frac{\nu }{2}\left( C-C^{\ast }\right) ,
\end{equation*}
with $C$ and $E$ called collapse and energy operators respectively. This
corresponds to the stochastic decoherence equation of the form 
\begin{equation*}
\mathrm{d}\chi \left( t\right) +\left( \frac{\nu }{2}\left( C^{\ast
}C-I\right) +\frac{i}{\hbar }E\right) \chi \left( t\right) \mathrm{d}%
t=\left( C-I\right) \chi \left( t\right) \mathrm{d}n_{t},
\end{equation*}
which was derived for quantum jumps caused by the counting observation in 
\cite{Be89a, BaBe}. It correspond to the linear stochastic decoherence
Master-equation 
\begin{equation*}
\mathrm{d}\varrho \left( t\right) +\left[ G\varrho \left( t\right) +\varrho
\left( t\right) G^{\ast }-\nu \varrho \left( t\right) \right] \mathrm{d}t=%
\left[ C\varrho \left( t\right) C^{\ast }-\varrho \left( t\right) \right] 
\mathrm{d}n_{t},\quad \varrho \left( 0\right) =\rho ,
\end{equation*}
for the not normalized (but normalized in the mean) density matrix $\varrho
\left( t,\omega \right) $, where $G=\frac{\nu }{2}C^{\ast }C+\frac{i}{\hbar }%
E$ (it has the form $\chi \left( t,\omega \right) \chi \left( t,\omega
\right) ^{\dagger }$ in the case of a pure initial state $\rho =\psi \psi
^{\dagger }$).

The nonlinear filtering equation for $\psi _{\omega }\left( t\right) $ in
this case has the form \cite{Be90a} 
\begin{equation*}
\mathrm{d}\psi _{\omega }+\left( \frac{\nu }{2}\left( C^{\ast }C-\left\|
C\psi _{\omega }\right\| ^{2}\right) +\frac{i}{\hbar }E\right) \psi _{\omega
}\mathrm{d}t=\left( C/\left\| C\psi _{\omega }\right\| -I\right) \psi
_{\omega }\mathrm{d}n_{\omega }^{\rho },
\end{equation*}
where $\left\| \psi \right\| =\left\langle \psi |\psi \right\rangle ^{1/2}$
(see also \cite{BeMe96} for the infinite-dimensional case). It corresponds
to the nonlinear stochastic Master-equation 
\begin{equation*}
\mathrm{d}\rho _{\omega }+\left[ G\rho _{\omega }+\rho _{\omega }G^{\ast
}-\nu \rho _{\omega }\mathrm{Tr}C\rho _{\omega }C^{\ast }\right] \mathrm{d}t=%
\left[ C\rho _{\omega }C^{\ast }/\mathrm{Tr}C\rho _{\omega }C^{\ast }-\rho %
\right] \mathrm{d}n_{\omega }^{\rho }
\end{equation*}
for the posterior density matrix $\rho _{\omega }\left( t\right) $ which is
the projector $\psi _{\omega }\left( t\right) \psi _{\omega }\left( t\right)
^{\dagger }$ for the pure initial state $\rho _{\omega }\left( 0\right)
=P_{\psi }$. Here $n^{\rho }\left( t\right) =n_{t}^{\rho \left( t\right) }$
is the output counting process which is described by the history probability
measure 
\begin{equation*}
\mathrm{P}\left( t,\mathrm{d}\omega \right) =\pi \left( t,\omega \right) \mu
\left( \mathrm{d}\omega \right) ,\quad \pi \left( t,\omega \right) =\mathrm{%
Tr}\varrho \left( t,\omega \right)
\end{equation*}
with the increment $\mathrm{d}n^{\rho }\left( t\right) $ independent of $%
n_{t}^{\rho }$ under the condition $\rho _{\omega }\left( t\right) =\rho $,
with the conditional expectation 
\begin{equation*}
\mathrm{M}\left[ \mathrm{d}n^{\rho }\left( t\right) |\rho _{\omega }\left(
t\right) =\rho \right] =\nu \mathrm{Tr}C\rho C^{\ast }\mathrm{d}t
\end{equation*}
(or $\nu \left\| C\psi \right\| ^{2}\mathrm{d}t$ for $\rho =P_{\psi }$). The
derivation and solution of this equation was also considered in \cite{BeSt91}%
, and its solution was applied in quantum optics in \cite{Car93, Car94}.

This nonlinear quantum jump equation can be written also in the quasi-linear
form \cite{Be89b, Be90a} 
\begin{equation*}
\mathrm{d}\psi_{\omega}\left( t\right) +\tilde{K}\left( t\right)
\psi_{\omega}\left( t\right) \mathrm{d}t=\tilde{L}\left( t\right)
\psi_{\omega}\left( t\right) \mathrm{d}y_{\omega}^{\rho}\left( t\right) ,
\end{equation*}
where $y_{\omega}^{\rho}\left( t\right) $ is the innovating martingale with
respect to the output measure which is described by the differential 
\begin{equation*}
\mathrm{d}y_{\omega}^{\rho}\left( t\right) =\nu^{-1/2}\left\| C\psi
_{\omega}\left( t\right) \right\| ^{-1}\mathrm{d}n_{\omega}^{\rho}\left(
t\right) -\nu^{1/2}\left\| C\psi_{\omega}\left( t\right) \right\| \mathrm{d}t
\end{equation*}
with $\rho=P_{\psi}$ and the initial $y_{\omega}^{\rho}\left( 0\right) =0 $,
the operator $\tilde{K}\left( t\right) $ similar to $K$ has the form 
\begin{equation*}
\tilde{K}\left( t\right) =\frac{1}{2}\tilde{L}\left( t\right) ^{\ast }\tilde{%
L}\left( t\right) +\frac{i}{\hbar}\tilde{H}\left( t\right) ,
\end{equation*}
and $\tilde{H}\left( t\right) ,\tilde{L}\left( t\right) $ depend on $t$ (and 
$\omega$) through the dependence on $\psi=\psi_{\omega}\left( t\right) $: 
\begin{equation*}
\tilde{L}=\nu^{1/2}\left( C-\left\| C\psi\right\| \right) ,\quad\tilde {H}%
=E+i\frac{\nu}{2}\left( C-C^{\ast}\right) \left\| C\psi\right\| .
\end{equation*}

The latter form of the nonlinear filtering equation admits the central limit 
$\nu \rightarrow \infty $ corresponding to the standard Wiener case $%
\varepsilon =0$ when $y_{t}=w_{t}$ with respect to the input Wiener measure $%
\mu $. If $L$ and $H$ do not depend on $\nu $, i.e. $C$ and $E$ depend on $%
\nu $ as 
\begin{equation*}
C=I+\nu ^{-1/2}L,\;E=H+\frac{\nu ^{1/2}}{2i}\left( L-L^{\ast }\right) ,
\end{equation*}
then $y_{\omega }^{\rho }\left( t\right) \rightarrow \tilde{y}_{t}$ as $%
\varepsilon ^{2}=\nu ^{-1}\rightarrow 0$, where the innovating diffusion
process $\tilde{y}_{t}$ defined as 
\begin{equation*}
\mathrm{d}\tilde{y}_{t}\left( \omega \right) =\mathrm{d}y_{t}\left( \omega
\right) -2\func{Re}\left\langle \psi _{\omega }\left( t\right) |L\psi
_{\omega }\left( t\right) \right\rangle \mathrm{d}t
\end{equation*}
can be identified with another standard Wiener process $\tilde{w}_{t}$ with
respect to the output probability measure due to $\tilde{\mu}\left( \mathrm{d%
}\omega \right) =\mu \left( \mathrm{d}\tilde{\omega}\right) $. If $\left\|
\psi _{\omega }\left( t\right) \right\| =1$ (which follows from the initial
condition $\left\| \psi \right\| =1$), the stochastic operator-functions $%
\tilde{L}\left( t\right) $, $\tilde{H}\left( t\right) $ defining the
nonlinear filtering equation have the limits 
\begin{equation*}
\tilde{L}=L-\func{Re}\left\langle \psi |L\psi \right\rangle ,\quad \tilde{H}%
=H+\frac{i}{2}\left( L-L^{\ast }\right) \func{Re}\left\langle \psi |L\psi
\right\rangle .
\end{equation*}
The corresponding nonlinear stochastic diffusion equation 
\begin{equation*}
\mathrm{d}\psi _{\omega }\left( t\right) +\tilde{K}\left( t\right) \psi
_{\omega }\left( t\right) \mathrm{d}t=\tilde{L}\left( t\right) \psi _{\omega
}\left( t\right) \mathrm{d}\tilde{w}_{t}
\end{equation*}
was first derived in the general multi-dimensional density-matrix form 
\begin{equation*}
\mathrm{d}\rho _{\omega }+\left[ K\rho _{\omega }+\rho _{\omega }K-L\rho
_{\omega }L^{\ast }\right] \mathrm{d}t=\left[ L\rho _{\omega }+\rho _{\omega
}L^{\ast }-\rho _{\omega }\mathrm{Tr}\left( L+L^{\ast }\right) \rho _{\omega
}\right] \mathrm{d}\tilde{w}_{t}
\end{equation*}
in \cite{Be88, Be90c} from the microscopic reversible quantum stochastic
evolution models by the quantum filtering method. It has been recently
applied in quantum optics \cite{WiMi93, GoGr93, WiMi94, GoGr94} for the
description of counting, homodyne and heterodyne time-continuous
measurements introduced in \cite{BaBe}. A particular case of this filtering
equation for the quantum particle in a potential field $\phi $ was also
derived phenomenologically by Diosi \cite{Dio88}. It was solved for the case
of linear and quadratic potential $\phi $ and the Gaussian initial wave
function in \cite{Be88, BeSt89}. The general microscopic derivation for the
case of multi-dimensional complete and incomplete measurements and solution
in the linear-Gaussian case is given in \cite{Be92b}. As in the classical
case this solution coincides with the optimal quantum linear filtering
(quantum Kalman filter) earlier obtained in \cite{Be80, Be85} for the
complex amplitude of the quantum open oscillator continuously observed in
the complex noise $y_{t}$ with $\left( \mathrm{d}y\right) ^{2}=0,\left| 
\mathrm{d}y\right| ^{2}=\mathrm{d}t$. The applications of this quantum
stochastic model to homodyne and heterodyne continuous measurements in
quantum optics can be found in \cite{GiPe93, WiMi94, GoGr94}. The explicit
solution of this stochastic wave equation for the free particle and the
Gaussian initial wave packet is given in \cite{BeSt92}. One can show \cite%
{ChSt92, Kol95} that the nondemolition observation of such a particle is
described by filtering of the quantum noise which results in the continual
collapse of any initial wave packet to the Gaussian stationary one localized
at the position posterior expectation.

The connection between the above diffusive nonlinear filtering equation and
our linear decoherence Master-equation 
\begin{equation*}
\mathrm{d}\varrho \left( t\right) +\left[ K\varrho \left( t\right) +\varrho
\left( t\right) K^{\ast }-L\varrho \left( t\right) L^{\ast }\right] \mathrm{d%
}t=\left[ L\varrho \left( t\right) +\varrho \left( t\right) L^{\ast }\right] 
\mathrm{d}w_{t},\quad \varrho \left( 0\right) =\rho ,
\end{equation*}
for the stochastic density operator $\varrho \left( t,\omega \right) $,
defining the output probability density $\mathrm{Tr}\varrho \left( t,\omega
\right) $, was well understood and presented in \cite{GPR90, GoGr94, GGH95}.
However it has also found an incorrect mathematical treatment in recent
Quantum State Diffusion theory \cite{Perc99} based on the case $\varepsilon
=0$ of our filtering equation (this particular nonlinear filtering equation
is empirically postulated as the `primary quantum state diffusion', and its
more fundamental linear version $\mathrm{d}\chi +K\chi \mathrm{d}t=L\chi 
\mathrm{d}w$ is `derived' in \cite{Perc99} simply by dropping the non-linear
terms without appropriate change of the probability measures for the
processes $\tilde{y}_{t}=\tilde{w}_{t}$ and $y_{t}=w_{t}$). The most general
stochastic decoherence Master equation is given in the Appendix 2.

\subsubsection{Q-bit trajectories and localizations}

Let us describe the exact Markovian model of an open quantum bit in a white
noise under nondemolition measurement. It was introduced in \cite{Be90b,
Be92a} even in the case of multidimensional quantum noise, but we shall
consider here just one dimension.

We assume that in an interaction representation picture the q-bit operators $%
S=\sigma \left( \mathbf{s}\right) $, $\mathbf{s}\in \mathbb{R}^{3}$ (e.g.
spins 1/2, see the notations of the Appendix 1) evolve as $S\left( t\right)
=U\left( t\right) ^{\ast }SU\left( t\right) $, where $U\left( t\right) $ is
stochastic unitary transformation in the Hilbert space $\mathfrak{h}=\mathbb{%
C}^{2}$ satisfying the It\^{o}-Schr\"{o}dinger equation 
\begin{equation*}
\mathrm{d}U\left( t\right) +\left( \frac{i}{\hbar }H+\frac{1}{2}L^{\ast
}L\right) U\left( t\right) \mathrm{d}t=\frac{i}{\hbar }LU\left( t\right) 
\mathrm{d}f_{t},\quad U\left( 0\right) =I.
\end{equation*}
Here $H=\sigma \left( \mathbf{h}\right) $ is a Hamiltonian and $L=\sigma
\left( \mathbf{l}\right) $ is a spin-operator given by real 3-vectors $%
\mathbf{h}$ and $\mathbf{l}$ (we assume for simplicity that $L^{\ast }=L$),
and $f_{t}=i\hbar \left( \Lambda _{-}-\Lambda ^{+}\right) _{t}$ is the
integral of the Langevin force which is defined as the input field momentum
process in the notations of the Appendix 2. The operators $S\left( t\right) $
satisfy the perturbed Heisenberg equation in the It\^{o} form 
\begin{equation*}
\mathrm{d}S\left( t\right) +\left( \frac{i}{\hbar }\left[ S\left( t\right)
,H\left( t\right) \right] +\frac{1}{2}\left[ \left[ S\left( t\right)
,L\left( t\right) \right] ,L\left( t\right) \right] \right) \mathrm{d}t=%
\frac{i}{\hbar }\left[ S\left( t\right) ,L\left( t\right) \right] \mathrm{d}%
f_{t}
\end{equation*}
and it can be written in the form of a vector stochastic equation 
\begin{equation*}
\mathrm{d}\mathbf{s}\left( t\right) +\left( \mathbf{k}\left( t\right) \times 
\mathbf{s}\left( t\right) +2\mathbf{l}\left( t\right) ^{2}\mathbf{s}\left(
t\right) -2\left( \mathbf{l}\left( t\right) \mathbf{\cdot s}\left( t\right)
\right) \mathbf{l}\left( t\right) \right) \mathrm{d}t=\frac{i}{\hslash }%
\left( \mathbf{l}\left( t\right) \times \mathbf{s}\left( t\right) \right) 
\mathrm{d}f_{t},
\end{equation*}
where $\mathbf{s}\left( 0\right) =\mathbf{s}$, $\mathbf{k}=2\hbar ^{-1}%
\mathbf{h}$, and time dependence of all coefficients is given in the
interaction representation of the corresponding operators $K=\sigma \left( 
\mathbf{k}\right) $, $L$.

Now we assume that indirect observation of this bit is the counting of the
output photon numbers $N_{t}=U\left( t\right) ^{\ast }\left( I\otimes
n_{t}\right) U\left( t\right) $, where 
\begin{equation*}
n_{t}=\nu t+\nu ^{1/2}\left( \Lambda _{-}+\Lambda ^{+}\right) _{t}+\Lambda
_{t}
\end{equation*}
is the Poisson process represented in Fock space as the quantum number
process of intensity $\nu $. One can easily prove by quantum It\^{o} formula
that this process, given by 
\begin{equation*}
\mathrm{d}N_{t}=\mathrm{d}\Lambda +\nu ^{1/2}\left( C\left( t\right) ^{\ast }%
\mathrm{d}\Lambda _{-}+C\left( t\right) \mathrm{d}\Lambda ^{+}\right) +\nu
C\left( t\right) ^{\ast }C\left( t\right) \mathrm{d}t,
\end{equation*}
where $C=I+\nu ^{-1/2}\sigma \left( \mathbf{l}\right) =C^{\ast }$, is
nondemolition so that the state of the q-bit can be predicted continuously
while it is measured in time. The optimal prediction in the mean square
sense of the q-bit operators $S\left( t\right) $ is given by the posterior
expectations 
\begin{equation*}
\left\langle S\right\rangle _{\omega }\left( t\right) =\mathrm{Tr}S\rho
_{\omega }\left( t\right) =\mathbf{s\cdot r}_{\omega }\left( t\right) ,
\end{equation*}
where $\mathbf{r}_{\omega }\left( t\right) $ is the posterior polarization
defining the posterior q-bit state 
\begin{equation*}
\rho _{\omega }\left( t\right) =\frac{1}{2}\left( \sigma \left( \mathbf{r}%
_{\omega }\left( t\right) \right) +I\right) .
\end{equation*}
It satisfies the nonlinear filtering equation \cite{Be90a, BaBe} 
\begin{align*}
& \mathrm{d}\mathbf{r}_{\omega }\left( t\right) +\left( 2\nu ^{1/2}\mathbf{l}%
+\mathbf{r}_{\omega }\left( t\right) \times \mathbf{k}-2\nu ^{1/2}\left( 
\mathbf{r}_{\omega }\left( t\right) \mathbf{\cdot l}\right) \mathbf{r}%
_{\omega }\left( t\right) \right) \mathrm{d}t \\
& =2\left( \frac{\nu ^{1/2}\left( \mathbf{l-}\left( \mathbf{r}_{\omega
}\left( t\right) \mathbf{\cdot l}\right) \mathbf{r}_{\omega }\left( t\right)
\right) +\left( \mathbf{l\cdot r}_{\omega }\left( t\right) \right) \mathbf{l}%
-\left( \mathbf{l\cdot l}\right) \mathbf{r}_{\omega }\left( t\right) }{\nu
+2\nu ^{1/2}\mathbf{l\cdot r}_{\omega }\left( t\right) +\mathbf{l\cdot l}}%
\right) \mathrm{d}n_{\omega }^{\rho }\left( t\right)
\end{align*}
with respect to the counting process $n_{\omega }^{\rho }\left( t\right) $
of the conditional intensity 
\begin{equation*}
\nu \mathrm{Tr}C\rho _{\omega }\left( t\right) C^{\ast }=\nu +2\nu ^{1/2}%
\mathbf{l}\cdot \mathbf{r}_{\omega }\left( t\right) +\mathbf{l\cdot l}.
\end{equation*}
The solution of this quantum filtering equation can be obtained in the form 
\begin{equation*}
\mathbf{r}_{\omega }\left( t\right) =\mathbf{p}\left( t,\omega \right) /\pi
\left( t,\omega \right) ,
\end{equation*}
where $\pi \left( t\right) $ is the probability density of the output
counting process $n_{\omega }^{\rho }\left( t\right) $ with respect to the
input probability measure for the Poisson process $n_{t}\left( \omega
\right) $, and $\mathbf{p}\left( t\right) $ is the polarization for the
stochastic density matrix $\varrho \left( t,\omega \right) $: 
\begin{equation*}
\pi \left( t\right) =\mathrm{Tr}\varrho \left( t\right) ,\quad \varrho
\left( t\right) =\frac{1}{2}\left( \sigma \left( \mathbf{p}\left( t\right)
\right) +\pi \left( t\right) I\right) .
\end{equation*}
The 4-vector $p=\left( \pi ,\mathbf{p}\right) $ is the solution of the
linear stochastic system 
\begin{equation*}
\mathrm{d}\pi \left( t\right) =\left( \nu ^{-1/2}\left( \mathbf{l\cdot l}%
\right) \pi \left( t\right) +2\mathbf{l\cdot p}\left( t\right) \right) 
\mathrm{d}y_{t},\quad \pi \left( 0\right) =1,
\end{equation*}
\begin{align*}
& \mathrm{d}\mathbf{p}\left( t\right) +\left( \mathbf{p}\left( t\right)
\times \mathbf{k+}2\left( \mathbf{l\cdot l}\right) \mathbf{p}\left( t\right)
-2\left( \mathbf{l\cdot p}\left( t\right) \right) \mathbf{l}\right) \mathrm{d%
}t \\
& =\left( 2\mathbf{l}\pi \left( t\right) +2\nu ^{-1/2}\left( \mathbf{l\cdot p%
}\left( t\right) \right) \mathbf{l}-\nu ^{-1/2}\left( \mathbf{l\cdot l}%
\right) \mathbf{p}\left( t\right) \right) \mathrm{d}y_{t},\quad \mathbf{p}%
\left( 0\right) =\mathbf{r,}
\end{align*}
where $y_{t}=\nu ^{-1/2}n_{t}-\nu ^{1/2}t$ is given by the stationary
Poisson process $n_{t}$ of the intensity $\nu $. The expectation 
\begin{equation*}
\mathbf{r}\left( t\right) =\int \mathbf{r}_{\omega }\left( t\right) \mathrm{P%
}\left( t,\mathrm{d}\omega \right) =\mathrm{M}\left[ \mathbf{p}\left(
t\right) \right]
\end{equation*}
gives the solution to the Bloch Master-equation 
\begin{equation*}
\frac{\mathrm{d}}{\mathrm{d}t}\mathbf{r}\left( t\right) +\mathbf{r}\left(
t\right) \times \mathbf{k+}2\left( \mathbf{l\cdot l}\right) \mathbf{r}\left(
t\right) =2\left( \mathbf{l\cdot r}\left( t\right) \right) \mathbf{l,\quad r}%
\left( 0\right) =\mathbf{r}
\end{equation*}
for the polarization of the averaged density matrix $\rho \left( t\right)
=\left( \sigma \left( \mathbf{r}\left( t\right) +I\right) \right) /2$.

Passing to the limit $\nu \rightarrow \infty $ of infinite intensity of the
counting nondemolition process we obtain the system of diffusive equations 
\begin{equation*}
\mathrm{d}\pi \left( t\right) =2\mathbf{l\cdot p}\left( t\right) \mathrm{d}%
w_{t},\quad \pi \left( 0\right) =1,
\end{equation*}
\begin{align*}
& \mathrm{d}\mathbf{p}\left( t\right) +\left( \mathbf{p}\left( t\right)
\times \mathbf{k+}2\left( \mathbf{l\cdot l}\right) \mathbf{p}\left( t\right)
-2\left( \mathbf{l\cdot p}\left( t\right) \right) \mathbf{l}\right) \mathrm{d%
}t \\
& =2\mathbf{l}\pi \left( t\right) \mathrm{d}w_{t},\quad \mathbf{p}\left(
0\right) =\mathbf{r,}
\end{align*}
where it is taken into account that the limit of the process $y_{t}$ is the
standard Wiener process $w_{t}$. This linear system, which was derived in 
\cite{Be90b}, represents the stochastic decoherence Master-equation for the
quantum bit under the continuous observation of the nondemolition process $%
Y_{t}=U\left( t\right) ^{\ast }\left( I\otimes w_{t}\right) U\left( t\right) 
$ given by 
\begin{equation*}
\mathrm{d}Y_{t}=\left( L\left( t\right) +L\left( t\right) ^{\ast }\right) 
\mathrm{d}t+\mathrm{d}\Lambda _{-}+\mathrm{d}\Lambda ^{+}=2L\left( t\right) 
\mathrm{d}t+\mathrm{d}w_{t},\quad Y_{0}=0,
\end{equation*}
Here $w_{t}=\left( \Lambda _{-}+\Lambda ^{+}\right) _{t}$ is the standard
Wiener process, defined as the input field coordinate process in Fock space
in the Appendix 2. It is the central limit of the Poisson process $n_{t}$ in
the Fock space, that is the limit of 
\begin{equation*}
y_{t}=\nu ^{-1/2}\Lambda _{t}+\Lambda _{-}+\Lambda ^{+}=\nu ^{-1/2}n_{t}-\nu
^{1/2}t
\end{equation*}
at $\nu \rightarrow \infty $. This limit quantum diffusion model for the
open quantum bit under the continuous observation coincides with the signal
+ noise model $Y\left( t\right) =\left| \mathbf{l}\right| X\left( t\right)
+e\left( t\right) $ considered for derivation of the generalized Heisenberg
principle. Here $\left| \mathbf{l}\right| =\left( \mathbf{l\cdot l}\right)
^{1/2}$, $X=\sigma \left( \mathbf{e}\right) $, $\mathbf{e}=\mathbf{l}/\left| 
\mathbf{l}\right| $ and $e\left( t\right) $ is the one half of the standard
white noise, the generalized derivative $w\left( t\right) =\mathrm{d}w_{t}/%
\mathrm{d}t.$ It was proved \ in \cite{Be88}, see also \cite{Be92b} for the
infinite dimensional case, that $Y_{t}$ is a commutative nondemolition
process with respect to the Heisenberg processes due to the canonical
commutation relations of $e\left( t\right) =w\left( t\right) /2$ and the
Langevin force $f\left( t\right) =\mathrm{d}f_{t}/\mathrm{d}t$. Note that
the quantum error process $w_{t}=2e_{t}$ does not commute with the
perturbing quantum process $f_{t}$ in Fock space due to the multiplication
table 
\begin{equation*}
\mathrm{d}f\mathrm{d}w=i\hbar \mathrm{d}t,\quad \mathrm{d}w\mathrm{d}%
f=-i\hbar \mathrm{d}t\text{.}
\end{equation*}
This corresponds to the canonical commutation relations for the normalized
derivatives $e\left( t\right) $ and $f\left( t\right) $ such that the true
Heisenberg principle is fulfilled at the boundary $\sigma \tau =\hbar /2$ of
the standard deviation $\sigma =1/2$ for $e$ and$\ \tau =\hbar $ for $f$.
Thus our quantum stochastic model of nondemolition observation is the
minimal perturbation model of the continual indirect measurement of the
quantum bit position $X\left( t\right) =\sigma \left( \mathbf{e}\left(
t\right) \right) $.

The solution of this linear diffusive system gives the solution $\mathbf{r}%
\left( t\right) =\mathbf{p}\left( t\right) /\pi \left( t\right) $ to the\
nonlinear filtering equation 
\begin{align*}
& \mathrm{d}\mathbf{r}_{\omega }\left( t\right) +\left( \mathbf{r}_{\omega
}\left( t\right) \times \mathbf{k+}2\left( \mathbf{l\cdot l}\right) \mathbf{r%
}_{\omega }\left( t\right) -2\left( \mathbf{l\cdot r}_{\omega }\left(
t\right) \right) \mathbf{l}\right) \mathrm{d}t \\
& =2\left( \mathbf{l-}\left( \mathbf{r}_{\omega }\left( t\right) \mathbf{%
\cdot l}\right) \mathbf{r}_{\omega }\left( t\right) \right) \mathrm{d}\tilde{%
w}_{t},\quad \mathbf{r}_{\omega }\left( 0\right) =\mathbf{r,}
\end{align*}
for the posterior diffusion of the quantum bit state under continuous
observation, derived in \cite{Be92a}. Here $\tilde{w}_{t}$ is an innovating
diffusive process given by the equation 
\begin{equation*}
\mathrm{d}\tilde{w}_{t}=\mathrm{d}w_{t}-2\mathbf{l\cdot r}_{\omega }\left(
t\right) \mathrm{d}t,\quad \tilde{w}_{0}=0.
\end{equation*}
It can be seen as another standard Wiener process, however not with respect
the input but the output probability measure corresponding to the histories
density $\pi \left( t,\omega \right) $. It is central limit of $n_{\omega
}^{\rho }\left( t\right) $ at $\nu \rightarrow \infty $, that is the limit
of the innovating counting martingale $y_{\omega }^{\rho }\left( t\right) $
given by the equation 
\begin{equation*}
\mathrm{d}y_{\omega }^{\rho }\left( t\right) =\nu ^{-1/2}\mathrm{d}n_{\omega
}^{\rho }\left( t\right) -\left( \nu ^{1/2}+2\mathbf{l}\cdot \mathbf{r}%
_{\omega }\left( t\right) +\nu ^{-1/2}\mathbf{l\cdot l}\right) \mathrm{d}t.
\end{equation*}

This solution can be easily obtained under the condition that $\mathbf{k}=k%
\mathbf{e}_{z}$ is colinear to $\mathbf{l}$, i.e. $\mathbf{e}=\mathbf{e}_{z}$%
, when this system splits into two independent systems 
\begin{equation*}
\mathrm{d}\pi\left( t\right) =2\left| \mathbf{l}\right| p_{z}\left( t\right) 
\mathrm{d}w_{t},\;\mathrm{d}p_{z}\left( t\right) =2\left| \mathbf{l}\right|
\pi\left( t\right) \mathrm{d}w_{t},
\end{equation*}
\begin{equation*}
\mathrm{d}\mathbf{p}_{\bot}\left( t\right) \mathbf{+}\left( 2\left| \mathbf{l%
}\right| ^{2}\mathbf{p}_{\bot}\left( t\right) +k\mathbf{p}_{\bot }\left(
t\right) \times\mathbf{e}_{z}\right) \mathrm{d}t=0,
\end{equation*}
where $p_{z}=\mathbf{e}_{z}\mathbf{\cdot p}$, $\mathbf{p}_{\bot}=\mathbf{p}%
-p_{z}\mathbf{e}_{z}$. The first stochastic system, diagonalized for $%
\pi_{\pm}=\left( \pi\pm p_{z}\right) /2$ as 
\begin{equation*}
\mathrm{d}\pi_{\pm}\left( t\right) =\pm2\left| \mathbf{l}\right| \pi_{\pm
}\left( t\right) \mathrm{d}w_{t},\quad\pi_{\pm}\left( 0\right) =\frac {1}{2}%
\left( 1\pm z\right) ,
\end{equation*}
has apparent solution $\pi=\pi_{-}+\pi_{+}$, $p_{l}=\pi_{+}-\pi_{-}$ where 
\begin{equation*}
\pi_{\pm}\left( t,\omega\right) =\frac{1}{2}\left( 1\pm z\right) \exp\left(
\pm2\left| \mathbf{l}\right| w_{t}-2\left| \mathbf{l}\right| ^{2}t\right)
\end{equation*}
are the joint propensity densities of the spin-projection $L=\sigma\left( 
\mathbf{l}\right) $ to be $\pm\left| \mathbf{l}\right| $ and the trajectory
of $Y$ to be $w$ up to the time $t$ with respect to the standard Wiener
probability measure $\mu$. This gives 
\begin{align*}
\pi\left( t,\omega\right) & =\left( \cosh2\left| \mathbf{l}\right|
w_{t}+z\sinh2\left| \mathbf{l}\right| w_{t}\right) \exp\left( -2\left| 
\mathbf{l}\right| ^{2}t\right) , \\
p_{z}\left( t,\omega\right) & =\left( \sinh2\left| \mathbf{l}\right|
w_{t}+z\cosh2\left| \mathbf{l}\right| w_{t}\right) \exp\left( -2\left| 
\mathbf{l}\right| ^{2}t\right) .
\end{align*}
The orthogonal component $\mathbf{p}_{\bot}$ has the spiral nonstochastic
rotation 
\begin{align*}
p_{x}\left( t\right) & =r_{x}^{0}\left( t\right) \exp\left( -2\left| \mathbf{%
l}\right| ^{2}t\right) ,\;r_{x}^{0}\left( t\right) =x\cos kt-y\sin kt, \\
p_{y}\left( t\right) & =r_{y}^{0}\left( t\right) \exp\left( -2\left| \mathbf{%
l}\right| ^{2}t\right) ,\;r_{y}^{0}\left( t\right) =y\cos kt+x\sin kt,
\end{align*}
which clearly suggests that $\mathbf{p}_{\bot}\left( t\right) \rightarrow0$
if $t\rightarrow\infty$ or $\left| \mathbf{l}\right| \rightarrow\infty$ for
each $t>0$.

Thus the stochastic components 
\begin{equation*}
\mathbf{r}_{\omega }^{\bot }\left( t\right) =\frac{\mathbf{p}_{\bot }\left(
t\right) }{\pi \left( t,\omega \right) }=x_{\omega }\left( t\right) \mathbf{e%
}_{x}+y_{\omega }\left( t\right) \mathbf{e}_{y},\quad z_{\omega }\left(
t\right) =\frac{p_{z}\left( t,\omega \right) }{\pi \left( t,\omega \right) }
\end{equation*}
of the posterior polarization $\mathbf{r}_{\omega }\left( t\right) $ are
found as 
\begin{align*}
\mathbf{r}_{\omega }^{\bot }\left( t\right) & =\left( \cosh 2\left| \mathbf{l%
}\right| w_{t}+z\sinh 2\left| \mathbf{l}\right| w_{t}\right) ^{-1}\mathbf{r}%
_{\bot }^{0}\left( t\right) \;, \\
z_{\omega }\left( t\right) & =\left( 1+z\tanh 2\left| \mathbf{l}\right|
w_{t}\right) ^{-1}\left( \tanh 2\left| \mathbf{l}\right| w_{t}+z\right) .
\end{align*}
Note that in order to express this as the solution to the nonlinear
filtering equation one has to make the substitution 
\begin{equation*}
\mathrm{d}w_{t}=2\mathbf{l\cdot r}_{\omega }\left( t\right) \mathrm{d}t+%
\mathrm{d}\tilde{w}_{t},\quad w_{0}=0.
\end{equation*}
At the limit $\left| \mathbf{l}\right| \rightarrow \infty $ of the infinite
accuracy of the nondemolition measurement $\mathbf{r}_{\omega }^{\bot
}\left( t\right) \rightarrow 0$, $z_{\omega }\left( t\right) \rightarrow 
\mathrm{sign}w_{t}=\pm 1$ for each finite $t$ and $w_{t}\neq 0$ independent
of the values $\mathbf{r}_{\bot }=x\mathbf{e}_{x}+y\mathbf{e}_{y}$ and $z$
for the components of the initial polarization $\mathbf{r}$. One can find
also the solution 
\begin{equation*}
\mathbf{r}\left( t\right) =\int \mathbf{p}\left( t,\omega \right) \mu \left( 
\mathrm{d}\omega \right) =\mathbf{p}_{\bot }\left( t\right) +z\mathbf{e}_{z}
\end{equation*}
to the Bloch Master-equation in the case of colinear $\mathbf{k}$ and $%
\mathbf{l}$.as the expectation of $\mathbf{r}_{\omega }\left( t\right) $
with respect to the output measure $\tilde{\mu}$. This apparently has the
limit $\mathbf{r}\left( t\right) \rightarrow z\mathbf{e}_{z}$ at $%
t\rightarrow \infty $.

Thus the stochastic decoherence, continuous trajectories and spontaneous
localizations of an open q-bit are derived as the result of continuous
nondemolition measurement of a spin-projection $\sigma\left( \mathbf{l}%
\right) $ plus white noise $e\left( t\right) $ from the unitary evolution
perturbed by another white noise $f\left( t\right) $.

\section{Conclusion: A quantum message from the future}

Although Schr\"{o}dinger didn't derive the stochastic filtering equation for
the continuously decohering wave function $\chi \left( t\right) $,
describing the state of the semiclassical system including the observable
nondemolition process $y_{t}$ in continuous time in the same way as we did
it for his cat just in one step, he did envisage a possibility of how to get
it `if one introduces two symmetric systems of waves, which are traveling in
opposite directions; one of them presumably has something to do with the
known (or supposed to be known) state of the system at a later point in
time' \cite{Schr31}. This desire coincides with the ``transactional'' \
attempt of interpretation of quantum mechanics suggested in \cite{Crm86} on
the basis that the relativistic wave equation yields in the nonrelativistic
limit two Schr\"{o}dinger type equations, one of which is the time reversed
version of the usual equation: `The state vector $\psi $ of the quantum
mechanical formalism is a real physical wave with spatial extension and it
is identical with the initial ``offer wave'' of the transaction. The
particle (photon, electron, etc.) and the collapsed state vector are
identical with the completed transaction.' \ There was no mathematical proof
of this statement in \cite{Crm86}, and it is obviously not true for the
deterministic state vector $\psi \left( t\right) $ satisfying the
conventional Schr\"{o}dinger equation, but we are going to show that this
interpretation is true for the stochastic wave $\chi \left( t\right) $
satisfying our decoherence equation.

First let us note that the stochastic equation for the offer wave $\chi
\left( t\right) $ and the standard input probability measure $\mu $ can be
represented in Fock space as 
\begin{equation*}
\mathrm{d}\chi \left( t\right) +K\chi \left( t\right) \mathrm{d}t=L\mathrm{d}%
y_{t}\chi \left( t\right) ,\quad \chi \left( 0\right) =\psi \otimes \delta
_{0},
\end{equation*}
where $y_{t}=\Lambda ^{+}+\Lambda _{-}+\varepsilon \Lambda $ in the notation
explained in the Appendix. It coincides on the noise vacuum state $\delta
_{0}$ with the quantum stochastic Schr\"{o}dinger equation 
\begin{equation*}
\mathrm{d}\varphi \left( t\right) +K\varphi \left( t\right) \mathrm{d}%
t=\left( L\mathrm{d}\Lambda ^{+}-L^{\ast }\mathrm{d}\Lambda _{-}\right)
\varphi \left( t\right) ,\quad \varphi \left( 0\right) =\psi \otimes \delta
_{0}
\end{equation*}
corresponding to the generalized Heisenberg equation with the Langevin
force, $if_{t}=\hbar \left( \Lambda ^{+}-\Lambda _{-}\right) _{t}$, if $%
L^{\ast }=L$. Indeed, as it was noted in \cite{Be92b}, due to adaptedness
both $L\mathrm{d}y$ and $L\mathrm{d}\Lambda ^{+}-L^{\ast }\mathrm{d}\Lambda
_{-}$ act on the tensor product states with future vacuum $\delta _{0}$ on
which they have the same action since $\Lambda _{-}\delta _{0}=0$, $\Lambda
\delta _{0}=0$ (the annihilation process $\Lambda _{-}$ is zero on the
vacuum $\delta _{0}$, as well as the number process $\Lambda $). Thus when
extended from $\delta _{0}$ to any initial Fock vector $\varphi _{0}$,
quantum stochastic evolution is the HP unitary propagation \cite{HuPa84}
which is a unitary cocycle on Fock space over $L^{2}\left( \mathbb{R}%
_{+}\right) $ with respect to the free time-shift evolution $\varphi \left(
t,s\right) =\varphi \left( 0,s+t\right) $ in the Fock space. This free plain
wave evolution in the half space $s>0$ in the extra dimension is the input,
or offer wave evolution for our three dimensional (or more?) world located
at the boundary of $\mathbb{R}_{+}$. The single offer waves do not interact
in the Fock space until they reach the boundary $s=0$ where they produce the
quantum jumps described by the stochastic differential equation.

As has been recently shown in \cite{Be00, Be01}, by doubling the Fock space
it is possible to extend the cocycle to a unitary group evolution which will
also include the free propagation of the output waves in the opposite
direction. The conservative boundary condition corresponding to the
interaction with our world at the boundary, includes the creation,
annihilation and exchange of the input-output waves. The corresponding ``Schr%
\"{o}dinger'' boundary value problem is the second quantization of the Dirac
wave equation on the half line, with a boundary condition in Fock space
which is responsible for the stochastic interaction of quantum noise with
our world in the course of the transaction of the input-output waves. These
nondemolition continual observations are represented in this picture by
measurement at the boundary of the arrival times and positions of the
particles corresponding to the quantized waves in Fock space with respect to
an ``offer state'', the input vacuum, dressed into the output wave. The
continual reduction process for our world wave function is then simply
represented as the decohering input wave function in the extended space,
which is filtered from the corresponding mixture of pure states by the
process of innovation of the initial knowledge during the continual
measurement. The result of this filtering gives the best possible prediction
of future states which is allowed by the quantum causality. As was shown on
the example of a free quantum particle under observation, the filtering
appears as a dissipation, oscillation and gravitation as a result of
nondemolition observation.

My friend Robin Hudson wrote in his \ Lecture Notes on Quantum Theory:

\begin{quote}
\textit{Quantum theory is a beautiful mathematical theory. If only it didn't
have to mean something, to be interpreted.}
\end{quote}

Obviously here he used beautiful in the sense of simple: Everything that is
simple is indeed beautiful. However Nature is beautiful but not simple: we
live at the edge of two worlds, one is quantum, the other one is classical,
everything in the future is quantized waves, everything in the past is
trajectories of recorded particles.

Certainly all great founders of quantum theory are followers of those about
whom Aristotle wrote in his Metaphysica:

\begin{quote}
\textit{they fancied that the principles of mathematics are the principles
of all things'}, and `...\textit{these are the greatest forms of
beauty.\medskip }
\end{quote}

\textsc{Appendix 1}\textbf{: On Bell's ``Proof'' that von Neumann's Proof
was in Error.}

To ``disprove'' the von Neumann's theorem on the nonexistence of hidden
variables in quantum mechanics Bell \cite{Bell66} argued that the
dispersion-free states specified by a hidden parameter $\lambda $ should be
additive only for commuting pairs from the space $\mathcal{L}=\mathcal{L}%
\left( \mathfrak{h}\right) $ of all Hermitian operators on the system
Hilbert space $\mathfrak{h}$. One can assume even less, that the
corresponding probability function $E\mapsto \left\langle E\right\rangle
_{\lambda }$ should be additive with respect to only orthogonal
decompositions in the subset $\mathcal{E}=\mathcal{E}\left( \mathfrak{h}%
\right) $ of all Hermitian projectors $E$, as only orthogonal events are
simultaneously verifiable by measuring an observable $L\in \mathcal{L}$. In
the case of finite-dimensional Hilbert space $\mathfrak{h}$ it is equivalent
to the Bell's assumption, but we shall reformulate his only counterexample
it terms of the propositions, or events $E\in \mathcal{E}$ in order to
dismiss his argument that he `is not dealing with logical propositions, but
with measurements involving, for example, differently oriented magnets' (p.6
in \cite{Bell87}).

Bell constructed an example of hidden dispersion-free states for the
quantum-mechanical states, described by one-dimensional projectors 
\begin{equation*}
\rho =\frac{1}{2}\left( I+\sigma \left( \mathbf{r}\right) \right) \equiv
P\left( \mathbf{r}\right) ,\quad \sigma \left( \mathbf{r}\right) =x\sigma
_{x}+y\sigma _{y}+z\sigma _{z}
\end{equation*}
in two-dimensional space $\mathfrak{h}=\mathbb{C}^{2}$, given by the points $%
\mathbf{r}=x\mathbf{e}_{x}+y\mathbf{e}_{y}+z\mathbf{e}_{z}$ on the unit
sphere $\mathbf{S}\subset \mathbb{R}^{3}$ and Pauli matrices $\sigma $. He
assigned to spin operators $\sigma \left( \mathbf{e}\right) $ describing the
spin projections in the directions $\mathbf{e\in S}$ the simultaneously
definite values 
\begin{equation*}
s_{\lambda }\left( \mathbf{e}\right) =\pm 1\equiv \left\langle \sigma \left( 
\mathbf{e}\right) \right\rangle _{\lambda },\quad \mathbf{e\in S}_{\lambda
}^{\pm }\left( \mathbf{r}\right) ,
\end{equation*}
which can be taken as their dispersion-free expectations $\left\langle
\sigma \left( \mathbf{e}\right) \right\rangle _{\lambda }$ due to 
\begin{equation*}
\sigma \left( -\mathbf{e}\right) =-\sigma \left( \mathbf{e}\right) ,\quad
\sigma \left( \mathbf{e}\right) ^{2}=I
\end{equation*}
and $\left\langle I\right\rangle _{\lambda }=1$ if $s_{\lambda }\left( -%
\mathbf{e}\right) =-s_{\lambda }\left( \mathbf{e}\right) $. This is
specified by a reflection-symmetric partition 
\begin{equation*}
\mathbf{S}_{\lambda }^{-}=-\mathbf{S}_{\lambda }^{+},\quad \mathbf{S}%
_{\lambda }^{-}\cup \mathbf{S}_{\lambda }^{+}=\mathbf{S,\quad S}_{\lambda
}^{-}\cap \mathbf{S}_{\lambda }^{+}=\emptyset
\end{equation*}
of the unit sphere $\mathbf{S}$. Obviously there are plenty of such
partitions, but Bell took a special family 
\begin{equation*}
\mathbf{S}_{\lambda }^{\pm }\left( \mathbf{r}\right) =\left[ \mathbf{S}^{\pm
}\left( \mathbf{r}\right) \backslash \mathbf{S}_{\lambda }\left( \pm \mathbf{%
r}\right) \right] \cup \left[ \mathbf{S}^{\mp }\left( \mathbf{r}\right)
\backslash \mathbf{S}_{-\lambda }\left( \pm \mathbf{r}\right) \right] ,
\end{equation*}
where $\mathbf{S}^{\pm }$ are south and north hemispheres of the standard
reflection-symmetric partition with $\mathbf{r}$ pointing north, and 
\begin{equation*}
\mathbf{S}_{\lambda }\left( \mathbf{r}\right) =\left\{ \mathbf{e}\in \mathbf{%
S}:\mathbf{e\cdot r}<2\lambda \right\}
\end{equation*}
is parametrized by $\lambda \in \left[ -\frac{1}{2},\frac{1}{2}\right] $ in
such a way that 
\begin{equation*}
\int_{-1/2}^{1/2}s_{\lambda }\left( \mathbf{e}\right) \mathrm{d}\lambda =%
\mathrm{\Pr }\left\{ \lambda :\mathbf{S}_{\lambda }^{+}\left( \mathbf{r}%
\right) \ni \mathbf{e}\right\} -\mathrm{\Pr }\left\{ \lambda :\mathbf{S}%
_{\lambda }^{-}\left( \mathbf{r}\right) \ni \mathbf{e}\right\} =\mathbf{%
e\cdot r.}
\end{equation*}
In his formula $\mathbf{r=e}_{z}$, but it can be extended to the case $%
\left| \mathbf{r}\right| \leq 1$ of not completely polarized quantum states $%
\rho $ defining the quantum-mechanical expectations $\left\langle \sigma
\left( \mathbf{e}\right) \right\rangle $ and quantum probabilities $\Pr
\left\{ P\left( \mathbf{e}\right) =1\right\} $ of the propositions $%
E=P\left( \mathbf{e}\right) $ as the linear and affine forms in the unit
ball of such $\mathbf{r}$: 
\begin{equation*}
\mathrm{Tr}\sigma \left( \mathbf{e}\right) \rho =\mathbf{e\cdot r,\quad }%
\mathrm{Tr}P\left( \mathbf{e}\right) \rho =\frac{1}{2}\left( 1+\mathbf{%
e\cdot r}\right) .
\end{equation*}
Each $\lambda $ assigns the zero-one probabilities $\left\langle P\left( \pm 
\mathbf{e}\right) \right\rangle _{\lambda }=\chi _{\mathbf{\lambda }}^{\pm
}\left( \mathbf{e}\right) $ given by the characteristic functions $\chi
_{\lambda }^{\pm }$ of $\mathbf{S}_{\lambda }^{\pm }$ simultaneously for all
quantum events $P\left( \pm \mathbf{e}\right) $, the eigen-projectors of $%
\sigma \left( \mathbf{e}\right) $ corresponding to the eigenvalues $\pm 1$: 
\begin{equation*}
P\left( \pm \mathbf{e}\right) =\frac{1}{2}\left( I\pm \sigma \left( \mathbf{e%
}\right) \right) \mapsto \chi _{\lambda }^{\pm }\left( \mathbf{e}\right) =%
\frac{1}{2}\left( 1\pm s_{\lambda }\left( \mathbf{e}\right) \right) \text{.}
\end{equation*}
The additivity of the probability function $E\mapsto \left\langle
E\right\rangle _{\lambda }$ in $\mathcal{E}=\left\{ O,P\left( \mathbf{S}%
\right) ,I\right\} $ at each $\lambda $ follows from $\left\langle
O\right\rangle _{\lambda }=0$: 
\begin{equation*}
\left\langle O\right\rangle _{\lambda }+\left\langle I\right\rangle
_{\lambda }=1=\left\langle O+I\right\rangle _{\lambda },
\end{equation*}
as $O+I=I$, and from $\chi _{\lambda }^{+}\left( -\mathbf{e}\right) =\chi
_{\lambda }^{-}\left( \mathbf{e}\right) $: 
\begin{equation*}
\left\langle P\left( \mathbf{e}\right) \right\rangle _{\lambda
}+\left\langle P\left( -\mathbf{e}\right) \right\rangle _{\lambda
}=1=\left\langle P\left( \mathbf{e}\right) +P\left( -\mathbf{e}\right)
\right\rangle _{\lambda },
\end{equation*}
as $P\left( \mathbf{e}\right) +P\left( -\mathbf{e}\right) =I$.

Thus a classical hidden variable theory reproducing the affine quantum
probabilities $\mathrm{P}\left( \mathbf{e}\right) =\left\langle P\left( 
\mathbf{e}\right) \right\rangle $ as the uniform mean value 
\begin{equation*}
\mathrm{M}\left\langle P\left( \mathbf{e}\right) \right\rangle _{\cdot
}=\int_{-1/2}^{1/2}\frac{1}{2}\left( 1+s_{\lambda }\left( \mathbf{e}\right)
\right) \mathrm{d}\lambda =\frac{1}{2}\left( 1+\mathbf{e\cdot r}\right) =%
\mathrm{Tr}P\left( \mathbf{e}\right) \rho
\end{equation*}
of the classical yes-no observables $\chi _{\cdot }^{+}\left( \mathbf{e}%
\right) =\left\langle P\left( \mathbf{e}\right) \right\rangle _{\cdot }$ was
constructed by Bell. However it does not contradict to the von Neumann
theorem even if the latter is strengthened by the restriction of the
additivity only to the orthogonal projectors $E\in \mathcal{E}$.

Indeed, apart from partial additivity (the sums are defined in $\mathcal{E}$
only for the orthogonal pairs from $\mathcal{E}$), the von Neumann theorem
restricted to $\mathcal{E}\subset \mathcal{L}$ should also inherit the \emph{%
physical continuity}, induced by ultra-strong topology in $\mathcal{L}$. In
the finite dimensional case it is just ordinary continuity in the projective
topology $\mathfrak{h}$, and in the case $\dim \mathfrak{h}=2$ it is the
continuity on the projective space $\mathbf{S}$ of all one-dimensional
projectors $P\left( \mathbf{e}\right) $, $\mathbf{e}\in \mathbf{S}$. It is
obvious that the zero-one probability function $E\mapsto \left\langle
E\right\rangle _{\lambda }$ constructed by Bell is not \emph{physically
continuous} on the restricted set: the characteristic function $\chi
_{\lambda }^{+}\left( \mathbf{e}\right) =\left\langle P\left( \mathbf{e}%
\right) \right\rangle _{\lambda }$ of the half-sphere $\mathbf{S}_{\lambda
}^{+}\left( \mathbf{r}\right) $ is discontinuous in $\mathbf{e}$ on the
whole sphere $\mathbf{S}$ for any $\lambda $ and $\mathbf{r}$. Measurements
of the spin projections in the physically close directions $\mathbf{e}%
_{n}\rightarrow \mathbf{e}$ should be described by close probabilities $%
\left\langle P\left( \mathbf{e}_{n}\right) \right\rangle _{\lambda
}\rightarrow \left\langle P\left( \mathbf{e}\right) \right\rangle _{\lambda }
$ in any physical state specified by $\lambda $, otherwise the state cannot
have physical meaning!

Moreover, the mean $\mathrm{M}$ over $\lambda$ cannot be considered as the
conditional averaging of a classical partially hidden world with respect to
the quantum observable part because it gives nonlinear expectations with
respect to the states $\rho$ even if it is restricted to the smallest
commutative algebra generated by the characteristic functions $\left\{
\chi_{\cdot}^{+}\left( \mathbf{e}\right) :\mathbf{e}\in\mathbf{S}\right\} $
of $\lambda:\mathbf{S}_{\lambda}^{+}\left( \mathbf{r}\right) \ni\mathbf{e}.$
One can see this by the uniform averaging of the commutative products $%
\chi_{\lambda}^{+}\left( \mathbf{e}\right) \chi_{\lambda}^{+}\left( \mathbf{f%
}\right) $: such mean values ( i.e. the second order moments) are affine
with respect to $\mathbf{r}$ only for colinear $\mathbf{e}$ and $\mathbf{f}%
\in\mathbf{S}$.

The continuity argument might be considered to be as purely mathematical,
but in fact it is not: even in classical probability theory with a discrete
phase space the pure states defined by Dirac $\delta $-measure, are
uniformly continuous, as any positive probability measure is on the space of
classical observables defined by bounded measurable functions on any
continuous phase space. In quantum theory an expectation defined as a linear
positive functional on $\mathcal{L}$ is also uniformly continuous, hence the
von Neumann assumption of physical (ultra-weak) continuity is only a
restriction in the infinite-dimensional case. Even if the state is defined
only on $\mathcal{E}\subset \mathcal{L}$ as a probability function which is
additive only on the orthogonal projectors, the uniform continuity follows
from its positivity in the case of $\dim \mathfrak{h}\geq 3$ \cite{Glea57}.
In fact, Gleason obtained more than this: He proved that the case $\dim 
\mathfrak{h}=2$ is the only exceptional one when a probability function on $%
\mathcal{E}\left( \mathfrak{h}\right) $ (which should be countably additive
in the case $\dim \mathfrak{h}=\infty $) may not be induced by a density
operator $\rho $, and thus cannot be extended to a linear expectation on the
operator space $\mathcal{L}\left( \mathfrak{h}\right) $. Such irregular
states cannot be extended by linearity on the algebra of all (not just
Hermitian) operators in $\mathfrak{h}=\mathbb{C}^{2}$ even if it is
continuous.

To rule out even this exceptional case we note that an irregular states $%
E\mapsto \left\langle E\right\rangle $ on $\mathcal{E}\left( \mathbb{C}%
^{2}\right) $ cannot be composed with any state of an additional quantum
system even if the latter is given by a regular probability function $%
\left\langle F\right\rangle =\mathrm{Tr}F\sigma $ on a set $\mathcal{E}%
\left( \mathfrak{g}\right) $ of ortho-projectors of another Hilbert space.
There is no additive probability function on the set $\mathcal{E}\left( 
\mathbb{C}^{2}\otimes \mathfrak{g}\right) $of all verifiable events for the
compound quantum system described by a nontrivial Hilbert space $\mathfrak{g}
$ such that 
\begin{equation*}
\left\langle E\right\rangle =\left\langle E\otimes I\right\rangle ,\quad
\left\langle I\otimes F\right\rangle =\mathrm{Tr}F\sigma ,
\end{equation*}
where $\sigma =P_{\varphi }$ is the density operator of wave function $%
\varphi \in \mathfrak{g}$. Indeed, if it could be possible for some $%
\mathfrak{g}$ with $\dim \mathfrak{g}>1$, it would be possible for $%
\mathfrak{g}=\mathbb{C}^{2}$. By virtue of Gleason's theorem any probability
function which is additive for orthogonal projectors on $\mathbb{C}%
^{2}\otimes \mathbb{C}^{2}=\mathbb{C}^{4}$ is regular on $\mathcal{E}\left( 
\mathbb{C}^{4}\right) $, given by a density operator $\hat{\varrho}$. Hence 
\begin{equation*}
\left\langle E\right\rangle =\mathrm{Tr}\left( I\otimes E\right) \hat{\varrho%
}=\mathrm{Tr}E\rho
\end{equation*}
i.e. the state on $\mathcal{E}\left( \mathbb{C}^{2}\right) $ is also
regular, with the density operator in $\mathfrak{h}=\mathbb{C}^{2}$ given by
the partial trace 
\begin{equation*}
\rho =\mathrm{Tr}\left[ \hat{\varrho}|\mathfrak{h}\right] =\mathrm{Tr}_{%
\mathfrak{g}}\hat{\varrho}.
\end{equation*}

In order to obtain an additive product-state on $\mathcal{E}\left( \mathbb{C}%
^{2}\otimes \mathfrak{g}\right) $ satisfying 
\begin{equation*}
\left\langle E\otimes F\right\rangle =\left\langle E\right\rangle \mathrm{Tr}%
FP_{\varphi },\quad E\in \mathcal{E}\left( \mathbb{C}^{2}\right) ,F\in 
\mathcal{E}\left( \mathfrak{g}\right)
\end{equation*}
for a finite-dimensional $\mathfrak{g}=\mathbb{C}^{n}$ with $n>1$ it
necessary to define the state as an expectation on the whole unit ball $%
\mathcal{B}_{1} $ of the algebra $\mathcal{B}$ of all (not just Hermitian)
operators in $\mathbb{C}^{2}$. Indeed, any one-dimensional Hermitian
projector in $\mathbb{C}^{2}\otimes \mathbb{C}^{n}=\mathbb{C}^{2n}$ can be
described as an $n\times n$-matrix $\mathbf{E}=\left[ A_{j}A_{i}^{\ast }%
\right] $ with $2\times 2$-entries $A_{j}\in \mathcal{B}_{1}\left( \mathbb{C}%
^{2}\right) $, $j=1,\ldots ,n$ satisfying the normalization condition 
\begin{equation*}
\sum_{j=1}^{n}A_{j}^{\ast }A_{j}=P\left( \mathbf{e}\right) =\frac{1}{2}%
\left( I+\sigma \left( \mathbf{e}\right) \right)
\end{equation*}
for some $\mathbf{e\in S}$. These entries have the form 
\begin{equation*}
A=\alpha P\left( \mathbf{e}\right) +aQ\left( \mathbf{e}_{\bot }\right)
,\quad Q\left( \mathbf{e}_{\bot }\right) =\frac{1}{2}\sigma \left( \mathbf{e}%
_{\bot }\right) ,
\end{equation*}
where $\mathbf{e}_{\bot }$ is an orthogonal complex vector such that 
\begin{equation*}
i\mathbf{e}_{\bot }\mathbf{\times e}=\mathbf{e}_{\bot }\quad \mathbf{e}%
_{\bot }^{\ast }\cdot \mathbf{e}_{\bot }=2,\quad i\mathbf{e}_{\bot }^{\ast
}\times \mathbf{e}_{\bot }=2\mathbf{e},
\end{equation*}
and $\sum \left( \left| \alpha _{j}^{2}\right| +\left| a_{j}^{2}\right|
\right) =1$ corresponding to $\mathrm{Tr}\mathbf{E}=1$. The matrix elements 
\begin{equation*}
A_{j}A_{i}^{\ast }=\alpha _{j}\alpha _{i}^{\ast }P\left( \mathbf{e}\right)
+a_{j}a_{i}^{\ast }P\left( -\mathbf{e}\right) +\alpha _{j}a_{i}^{\ast
}Q\left( \mathbf{e}_{\bot }^{\ast }\right) +a_{j}\alpha _{i}^{\ast }Q\left( 
\mathbf{e}_{\bot }\right)
\end{equation*}
for these orthoprojectors in $\mathbb{C}^{2n}$ are any matrices from the
unit ball $\mathcal{B}_{1}\left( \mathbb{C}^{2}\right) $, not just Hermitian
orthoprojectors. By virtue of Gleason's theorem the product-state of such
events $\mathbf{E}$ must be defined by the additive probability 
\begin{equation*}
\left\langle \mathbf{E}\right\rangle =\sum_{i,j=1}^{n}\varphi ^{j}\left(
A_{j}A_{\iota }^{\ast }\right) \bar{\varphi}^{i}=\varrho \left( B\right) ,
\end{equation*}
where $B=A\left( \varphi \right) A\left( \varphi \right) ^{\ast }=\beta
I+\sigma \left( \mathbf{b}\right) $ is given by $\alpha \left( \varphi
\right) =\varphi ^{j}\alpha _{j}$, $a\left( \varphi \right) =\varphi
^{j}a_{j}$ for $\varphi \in \mathbb{C}^{n}$ with the components $\varphi
^{j}=\bar{\varphi}_{j}$, and $\varrho \left( B\right) =\mathrm{Tr}B\rho $ is
the linear expectation 
\begin{equation*}
\varrho \left( B\right) =\frac{1}{2}\left( \beta _{+}\left( 1+r_{1}\right)
+\beta _{-}\left( 1-r_{1}\right) +b_{\bot }\bar{r}_{\bot }+\bar{b}_{\bot
}r_{\bot }\right) =\beta +\mathbf{b\cdot r}
\end{equation*}
with $r_{1}=\mathbf{e\cdot r}$, $\mathbf{r}_{\bot }=\mathbf{e}_{\bot }%
\mathbf{\cdot r}$, $\beta _{+}=\left| \alpha \left( \varphi \right) \right|
^{2}$, $\beta _{-}=\left| a\left( \varphi \right) \right| ^{2}$, $b_{\bot
}=\alpha \left( \varphi \right) \overline{a\left( \varphi \right) }$. It
these terms we can formulate the definition of a regular state without
assuming a priori the linearity and even continuity conditions also for the
case $\mathfrak{h}=\mathbb{C}^{2}$.

A complex-valued map $B\mapsto \varrho \left( B\right) $ on the unit ball $%
\mathcal{B}_{1}\left( \mathfrak{h}\right) $ normalized as $\varrho \left(
I\right) =1$ is called\emph{\ state} for a quantum system described by the
Hilbert space $\mathfrak{h}$ (including the case $\dim \mathfrak{h}=2$) if
it is positive on all Hermitian projective matrices $\mathbf{E}=\left[
A_{j}A_{k}^{\ast }\right] $ with entries $A_{j}\in \mathcal{B}_{1}\left( 
\mathfrak{h}\right) $ in the sense 
\begin{equation*}
\sum_{j}A_{j}^{\ast }A_{j}=P\in \mathcal{E}\left( \mathfrak{h}\right)
\Rightarrow \varrho \left( \mathbf{E}\right) =\left[ \varrho \left(
A_{j}A_{k}^{\ast }\right) \right] \geq 0,
\end{equation*}
of positive-definiteness of the matrices $\varrho \left( \mathbf{E}\right) $
with the complex entries $\left[ \varrho \left( A_{j}A_{k}^{\ast }\right) %
\right] .$ It is called a \emph{regular state} if 
\begin{equation*}
\varrho \left( E\otimes P_{\varphi }\right) =\varrho \left( E\right)
P_{\varphi }
\end{equation*}
for any one-dimensional projector $P_{\varphi }=\left[ \varphi _{j}\varphi
_{i}^{\ast }\right] $, and\ if it is countably-additive with respect to the
orthogonal decompositions $\mathbf{E}=\sum \mathbf{E}\left( k\right) $: 
\begin{equation*}
\sum_{j}A_{j}\left( i\right) ^{\ast }A_{j}\left( k\right) =0,\forall i\neq
k\Rightarrow \varrho \left( \sum_{k}A_{j}\left( k\right) A_{j}\left(
k\right) ^{\ast }\right) =\sum_{k}\varrho \left( A_{j}\left( k\right)
A_{i}\left( k\right) ^{\ast }\right) .
\end{equation*}

It obvious that the state thus defined can be uniquely extended to a regular
product-state on $\mathcal{E}\left( \mathfrak{h}\otimes \mathbb{C}%
^{n}\right) $ by 
\begin{equation*}
\sum_{j,k}\bar{\varphi}_{j}\varrho \left( A_{j}A_{i}^{\ast }\right) \varphi
_{i}\geq 0,\quad \forall \varphi _{j}\in \mathbb{C},\quad \sum \left|
\varphi _{j}\right| =1,
\end{equation*}
which proves that it is continuous and is given by a density operator: $%
\varrho \left( B\right) =\mathrm{Tr}B\rho $. Thus the composition principle
rules out the existence of the hidden variable representation for the
quantum bits corresponding to the case $\mathfrak{h}=\mathbb{C}^{2}$.\medskip

\textsc{Appendix 2}\textbf{: Symbolic Calculus for Quantum Noise.}

In order to formulate the differential nondemolition causality condition and
to derive a filtering equation for the posterior states in the
time-continuous case we need quantum stochastic calculus.

The classical differential calculus for the infinitesimal increments 
\begin{equation*}
\mathrm{d}x=x\left( t+\mathrm{d}t\right) -x\left( t\right) 
\end{equation*}%
became generally accepted only after Newton gave a simple algebraic rule $%
\left( \mathrm{d}t\right) ^{2}=0$ for the formal computations of the
differentials $\mathrm{d}x$ for smooth trajectories $t\mapsto x\left(
t\right) $. In the complex plane $\mathbb{C}$ of phase space it can be
represented by a one-dimensional algebra $\mathfrak{a}=\mathbb{C}\mathrm{d}%
_{t}$ of the elements $a=\alpha \mathrm{d}_{t}$ with involution $a^{\star }=%
\bar{\alpha}\mathrm{d}_{t}$. Here 
\begin{equation*}
\text{$\mathrm{d}_{t}$}=\left[ 
\begin{array}{ll}
0 & 1 \\ 
0 & 0%
\end{array}%
\right] =\frac{1}{2}\left( \sigma _{1}+i\sigma _{2}\right) 
\end{equation*}%
for $\mathrm{d}t$ is the nilpotent matrix, which can be regarded as
Hermitian $\mathrm{d}_{t}^{\star }=\mathrm{d}_{t}$ with respect to the
Minkowski metrics $\left( \mathbf{z}|\mathbf{z}\right) =2\func{Re}z_{-}\bar{z%
}_{+}$ in $\mathbb{C}^{2}$.

This formal rule was generalized to non-smooth paths early in the last
century in order to include the calculus of forward differentials $\mathrm{d}%
w\simeq\left( \mathrm{d}t\right) ^{1/2}$ for continuous diffusions $w_{t}$
which have no derivative at any $t$, and the forward differentials $\mathrm{d%
}n\in\left\{ 0,1\right\} $ for left continuous counting trajectories $n_{t}$
which have zero derivative for almost all $t$ (except the points of
discontinuity where $\mathrm{d}n=1$). The first is usually done by adding
the rules 
\begin{equation*}
\left( \mathrm{d}w\right) ^{2}=\mathrm{d}t,\quad\mathrm{d}w\mathrm{d}t=0=%
\mathrm{d}t\mathrm{d}w
\end{equation*}
in formal computations of continuous trajectories having the first order
forward differentials $\mathrm{d}x=\alpha\mathrm{d}t+\beta\mathrm{d}w$ with
the diffusive part given by the increments of standard Brownian paths $w $.
The second can be done by adding the rules 
\begin{equation*}
\left( \mathrm{d}n\right) ^{2}=\mathrm{d}n,\quad\mathrm{d}n\mathrm{d}t=0=%
\mathrm{d}t\mathrm{d}n
\end{equation*}
in formal computations of left continuous and smooth for almost all $t$
trajectories having the forward differentials $\mathrm{d}x=\alpha \mathrm{d}%
t+\gamma\mathrm{d}m$ with jumping part given by the increments of standard
compensated Poisson paths $m_{t}=n_{t}-t$. These rules were developed by It%
\^{o} \cite{Ito51} into the form of a stochastic calculus.

The linear span of $\mathrm{d}t$ and $\mathrm{d}w$ forms the Wiener-It\^{o}
algebra $\mathfrak{b}=\mathbb{C}\mathrm{d}_{t}+\mathbb{C}\mathrm{d}_{w}$,
while the linear span of $\mathrm{d}t$ and $\mathrm{d}n$ forms the Poisson-It%
\^{o} algebra $\mathfrak{c}=\mathbb{C}\mathrm{d}_{t}+\mathbb{C}\mathrm{d}_{m}
$, with the second order nilpotent $\mathrm{d}_{w}=\mathrm{d}_{w}^{\star }$
and the idempotent $\mathrm{d}_{m}=\mathrm{d}_{m}^{\star }$. They are
represented together with $\mathrm{d}_{t}$ by the triangular Hermitian
matrices 
\begin{equation*}
\text{$\mathrm{d}_{t}$}=\left[ 
\begin{array}{lll}
0 & 0 & 1 \\ 
0 & 0 & 0 \\ 
0 & 0 & 0%
\end{array}%
\right] ,\quad \mathrm{d}_{w}=\left[ 
\begin{array}{lll}
0 & 1 & 0 \\ 
0 & 0 & 1 \\ 
0 & 0 & 0%
\end{array}%
\right] ,\emph{\quad }\mathrm{d}_{m}\mathbf{=}\left[ 
\begin{array}{lll}
0 & 1 & 0 \\ 
0 & 1 & 1 \\ 
0 & 0 & 0%
\end{array}%
\right] ,
\end{equation*}%
on the Minkowski space $\mathbb{C}^{3}$ with respect to the inner Minkowski
product $\left( \mathbf{z}|\mathbf{z}\right) =z_{-}z^{-}+z_{\circ }z^{\circ
}+z_{+}z^{+}$, where $z^{\mu }=\bar{z}_{-\mu }$, $-\left( -,\circ ,+\right)
=\left( +,\circ ,-\right) $.

Although both algebras $\mathfrak{b}$ and $\mathfrak{c}$ are commutative,
the matrix algebra $\mathfrak{a}$ generated by $\mathfrak{b}$ and $\mathfrak{%
c}$ on $\mathbb{C}^{3}$ is not: 
\begin{equation*}
\mathrm{d}_{w}\mathrm{d}_{m}=\left[ 
\begin{array}{lll}
0 & 1 & 1 \\ 
0 & 0 & 0 \\ 
0 & 0 & 0%
\end{array}%
\right] \neq \left[ 
\begin{array}{lll}
0 & 0 & 1 \\ 
0 & 0 & 1 \\ 
0 & 0 & 0%
\end{array}%
\right] =\mathrm{d}_{m}\mathrm{d}_{w}.
\end{equation*}%
The four-dimensional $\star $-algebra $\mathfrak{a}=\mathbb{C}\mathrm{d}_{t}+%
\mathbb{C}\mathrm{d}_{-}+\mathbb{C}\mathrm{d}^{+}+\mathbb{C}\mathrm{d}$ of
triangular matrices with the canonical basis 
\begin{equation*}
\mathrm{d}_{-}=\left[ 
\begin{array}{lll}
0 & 1 & 0 \\ 
0 & 0 & 0 \\ 
0 & 0 & 0%
\end{array}%
\right] ,\,\mathrm{d}^{+}\mathbf{=}\left[ 
\begin{array}{lll}
0 & 0 & 0 \\ 
0 & 0 & 1 \\ 
0 & 0 & 0%
\end{array}%
\right] ,\,\mathrm{d}=\left[ 
\begin{array}{lll}
0 & 0 & 0 \\ 
0 & 1 & 0 \\ 
0 & 0 & 0%
\end{array}%
\right] ,
\end{equation*}%
given by the algebraic combinations 
\begin{equation*}
\mathrm{d}_{-}=\mathrm{d}_{w}\mathrm{d}_{m}-\text{$\mathrm{d}_{t}$},\;%
\mathrm{d}^{+}=\mathrm{d}_{m}\mathrm{d}_{w}-\text{$\mathrm{d}_{t}$},\;\text{%
\textrm{d}}=\text{\textrm{d}}_{m}-\text{\textrm{d}}_{w}
\end{equation*}%
is the canonical representation of the differential $\star $-algebra for
one-dimensional vacuum noise in the unified quantum stochastic calculus \cite%
{Be88a, Be92a}. It realizes the HP (Hudson-Parthasarathy) table \cite{HuPa84}
\begin{equation*}
\mathrm{d}\Lambda _{-}\mathrm{d}\Lambda ^{+}=\mathrm{d}t,\quad \mathrm{d}%
\Lambda _{-}\mathrm{d}\Lambda =\mathrm{d}\Lambda _{-},\quad \mathrm{d}%
\Lambda \mathrm{d}\Lambda ^{+}=\text{\textrm{d}}\Lambda ^{+},\quad \left( 
\mathrm{d}\Lambda \right) ^{2}=\mathrm{d}\Lambda ,
\end{equation*}%
with zero products for all other pairs, for the multiplication of the
canonical counting $\mathrm{d}\Lambda =\lambda \left( \mathrm{d}\right) $,
creation $\mathrm{d}\Lambda ^{+}=\lambda \left( \mathrm{d}^{+}\right) $,
annihilation $\mathrm{d}\Lambda _{-}=\lambda \left( \mathrm{d}_{-}\right) $,
and preservation $\mathrm{d}t=\lambda \left( \text{$\mathrm{d}_{t}$}\right) $
quantum stochastic integrators in Fock space over $L^{2}\left( \mathbb{R}%
_{+}\right) $. As was proved recently in \cite{Be98}, any generalized It\^{o}
algebra describing a quantum noise can be represented in the canonical way
as a $\star $-subalgebra of a quantum vacuum algebra 
\begin{equation*}
\mathrm{d}\Lambda _{\mu }^{\iota }\mathrm{d}\Lambda _{\kappa }^{\nu }=\delta
_{\kappa }^{\iota }\mathrm{d}\Lambda _{\mu }^{\nu },\quad \kappa ,\mu \in
\left\{ -,1,\ldots ,d\right\} ;\;\iota ,\nu \in \left\{ 1,\ldots
,d,+\right\} ,
\end{equation*}%
in the Fock space with several degrees of freedom $d$, where $\mathrm{d}%
\Lambda _{-}^{+}=\mathrm{d}t$ and $d$ is restricted by the doubled
dimensionality of quantum noise (could be infinite), similar to the
representation of every semi-classical system with a given state as a
subsystem of quantum system with a pure state. Note that in this quantum It%
\^{o} product formula $\delta _{\kappa }^{\iota }=0$ if $\iota =+$ or $%
\kappa =-$ as $\delta _{\kappa }^{\iota }\neq 0$ only when $\iota =\kappa $.

The quantum It\^{o} product gives an explicit form 
\begin{equation*}
\mathrm{d}\chi \chi ^{\dagger }+\chi \mathrm{d}\chi ^{\dagger }+\mathrm{d}%
\chi \mathrm{d}\chi ^{\dagger }=\left( \alpha _{\nu }^{\mu }\chi ^{\dagger
}+\chi \alpha _{\nu }^{\star \mu }+\alpha _{k}^{\mu }a_{\nu }^{\star
k}\right) _{\nu }^{\mu }\mathrm{d}\Lambda _{\mu }^{\nu }
\end{equation*}
of the term $\mathrm{d}\chi \mathrm{d}\chi ^{\dagger }$ for the adjoint
quantum stochastic differentials 
\begin{equation*}
\mathrm{d}\chi =\alpha _{\nu }^{\mu }\mathrm{d}\Lambda _{\mu }^{\nu },\quad 
\mathrm{d}\chi ^{\dagger }=\alpha _{\nu }^{\star \mu }\mathrm{d}\Lambda
_{\mu }^{\nu },
\end{equation*}
for evaluation of the product differential 
\begin{equation*}
\mathrm{d}\left( \chi \chi ^{\dagger }\right) =\left( \chi +\mathrm{d}\chi
\right) \left( \chi +\mathrm{d}\chi \right) ^{\dagger }-\chi \chi ^{\dagger
}.
\end{equation*}
Here $\alpha _{-\nu }^{\star \mu }=\alpha _{-\mu }^{\nu \dagger }$ is the
quantum It\^{o} involution with respect to the switch $-\left( -,+\right)
=\left( +,-\right) $, $-\left( 1,\ldots ,d\right) =\left( 1,\ldots ,d\right) 
$, introduced in \cite{Be88a}, and the Einstein summation is always
understood over $\nu =1,\ldots ,d,+$; $\mu =-,1,\ldots ,d$ and $k=1,\ldots
,d $. This is the universal It\^{o} product formula which lies in the heart
of the general quantum stochastic calculus \cite{Be88a, Be92a} unifying the
It\^{o} classical stochastic calculi with respect to the Wiener and Poisson
noises and the quantum differential calculi \cite{HuPa84, GaCo85} based on
the particular types of quantum It\^{o} algebras for the vacuum or finite
temperature noises. It was also extended to the form of quantum functional It%
\^{o} formula and even for the quantum nonadapted case in \cite{Be91, Be93}.

In particular, any real-valued process $y_{t}$ with zero mean value $%
\left\langle y_{t}\right\rangle =0$ and independent increments generating a
two-dimensional It\^{o} algebra has the differential $\mathrm{d}y$ in the
form of a commutative combination of $\mathrm{d}\Lambda,\mathrm{d}\Lambda
_{-},\mathrm{d}\Lambda^{+}$. The It\^{o} formula for the process $y_{t}$ can
be obtained from the HP product 
\begin{equation*}
\mathrm{d}\chi\mathrm{d}\chi^{\dagger}=\alpha\alpha^{\dagger}\mathrm{d}%
\Lambda+\alpha^{-}\alpha^{\dagger}\mathrm{d}\Lambda_{-}+\alpha\alpha
^{-\dagger}\mathrm{d}\Lambda^{+}+\alpha^{-}\alpha^{-\dagger}\mathrm{d}t
\end{equation*}
for the quantum stochastic differential 
\begin{equation*}
\mathrm{d}\chi=\alpha\mathrm{d}\Lambda+\alpha^{-}\mathrm{d}%
\Lambda_{-}+\alpha_{+}\mathrm{d}\Lambda^{+}\mathrm{d}\chi+\alpha_{+}^{-}%
\mathrm{d}t.
\end{equation*}

The noise $y_{t}$ is called standard if it has stationary increments with
the standard variance $\left\langle y_{t}^{2}\right\rangle =t$. In this case 
\begin{equation*}
y_{t}=\left( \Lambda ^{+}+\Lambda _{-}+\varepsilon \Lambda \right)
_{t}=\varepsilon m_{t}+\left( 1-\varepsilon \right) w_{t},
\end{equation*}
where $\varepsilon \geq 0$ is defined by the equation $\left( \mathrm{d}%
y\right) ^{2}-\mathrm{d}t=\varepsilon \mathrm{d}y$. Such, and indeed higher
dimensional, quantum noises for continual measurements in quantum optics
were considered in \cite{GPZ92, DPZG92}.

The general form of a quantum stochastic decoherence equation, based on the
canonical representation of the arbitrary It\^{o} algebra for a quantum
noise in the vacuum of $d$ degrees of freedom, can be written as 
\begin{equation*}
\mathrm{d}\chi \left( t\right) =\left( L_{\nu }^{\mu }-\delta _{\nu }^{\mu
}\right) \chi \left( t\right) \mathrm{d}\Lambda _{\mu }^{\nu },\quad \chi
\left( 0\right) =\psi .
\end{equation*}
Here $L_{\nu }^{\mu }$ are the operators in the system Hilbert space $%
\mathfrak{h}$ $\ni \psi $ with $L_{\kappa }^{\star -}L_{+}^{\kappa }=0$ for
the mean square normalization 
\begin{equation*}
\left\langle \chi \left( t\right) ^{\dagger }\chi \left( t\right)
\right\rangle =\mathrm{M}\chi \left( t\right) ^{\dagger }\chi \left(
t\right) =\psi ^{\dagger }\psi
\end{equation*}
with respect to the vacuum of Fock space of the quantum noise, where the
Einstein summation is understood over all $\kappa =-,1,\ldots ,d,+$ with the
agreement 
\begin{equation*}
L_{-}^{-}=I=L_{+}^{+},\quad \,L_{-}^{k}=0=L_{k}^{+},\quad k=1,\ldots ,d
\end{equation*}
and $\delta _{\nu }^{\mu }=1$ for all coinciding $\mu ,\nu \in \left\{
-,1,\ldots ,d,+\right\} $ such that $L_{\nu }^{\mu }-\delta _{\nu }^{\mu }=0$
whenever $\mu =+$ or $\nu =-$. In the notations $L_{+}^{i}=L^{i}$, $%
L_{+}^{-}=-K$, $L_{j}^{-}=-K_{j}$, $i,j=1,\ldots ,d$ the decoherence wave
equation takes the standard form \cite{Be95, Be97} 
\begin{equation*}
\mathrm{d}\chi \left( t\right) +\left( K\mathrm{d}t+K_{j}\mathrm{d}\Lambda
_{-}^{j}\right) \chi \left( t\right) =\left( L^{i}\mathrm{d}\Lambda
_{i}^{+}+\left( L_{j}^{i}-\delta _{j}^{i}\right) \mathrm{d}\Lambda
_{i}^{j}\right) \chi \left( t\right) ,
\end{equation*}
where $\Lambda _{i}^{+}\left( t\right) ,\Lambda _{-}^{j}\left( t\right)
,\Lambda _{i}^{j}\left( t\right) $ are the canonical creation, annihilation
and exchange processes respectively in Fock space, and the normalization
condition is written as $L_{k}L^{k}=K+K^{\ast }$ with $L_{k}^{\ast }=L^{k}$
(the Einstein summation is over $i,j,k=1,\ldots ,d$).

Using the quantum It\^{o} formula one can obtain the corresponding equation
for the quantum stochastic density operator $\hat{\varrho}=\chi \chi
^{\dagger }$ which is the particular case $\kappa =-,1,\ldots ,d,+$ of the
general quantum stochastic Master equation 
\begin{equation*}
\mathrm{d}\hat{\varrho}\left( t\right) =\left( L_{\nu }^{\kappa }\hat{\varrho%
}\left( t\right) L_{\kappa }^{\star \mu }-\hat{\varrho}\left( t\right)
\delta _{\nu }^{\mu }\right) \mathrm{d}\Lambda _{\mu }^{\nu },\quad \hat{%
\varrho}\left( 0\right) =\rho ,
\end{equation*}
where the summation over $\kappa =-,k,+$ is extended to infinite number of $%
k=1,2,\ldots $. This general form of the decoherence equation with $%
L_{\kappa }^{\star -}L_{+}^{\kappa }=0$ corresponding to the normalization
condition $\left\langle \hat{\varrho}\left( t\right) \right\rangle =\mathrm{%
Tr}\rho $ in the vacuum mean, was recently derived in terms of quantum
stochastic completely positive maps in \cite{Be95, Be97}. Denoting $L_{\nu
}^{-}=-K_{\nu }$, $L_{+}^{\star \mu }=-K^{\mu }$ such that $K_{\mu }^{\ast
}=K^{\mu }$, this can be written as 
\begin{equation*}
\mathrm{d}\hat{\varrho}\left( t\right) +K_{\nu }\hat{\varrho}\left( t\right) 
\mathrm{d}\Lambda _{-}^{\nu }+\hat{\varrho}\left( t\right) K^{\mu }\mathrm{d}%
\Lambda _{\mu }^{+}=\left( L_{\nu }^{k}\hat{\varrho}\left( t\right)
L_{k}^{\star \mu }-\hat{\varrho}\left( t\right) \delta _{\nu }^{\mu }\right) 
\mathrm{d}\Lambda _{\mu }^{\nu },
\end{equation*}
or in the notation above, $K_{+}=K,K^{-}=K^{\ast }$, $L_{+}^{k}=L^{k}$, $%
L_{k}^{\star -}=L_{k}$, $L_{k}^{\star i}=L_{i}^{k\ast }$ as 
\begin{equation*}
\mathrm{d}\hat{\varrho}\left( t\right) +\left( K\hat{\varrho}\left( t\right)
+\hat{\varrho}\left( t\right) K^{\ast }-L^{k}\hat{\varrho}\left( t\right)
L_{k}\right) \mathrm{d}t=\left( L_{j}^{k}\hat{\varrho}\left( t\right)
L_{k}^{\ast i}-\hat{\varrho}\left( t\right) \delta _{j}^{i}\right) \mathrm{d}%
\Lambda _{i}^{j}
\end{equation*}
\begin{equation*}
+\left( L_{j}^{k}\hat{\varrho}\left( t\right) L_{k}-K_{j}\hat{\varrho}\left(
t\right) \right) \mathrm{d}\Lambda _{-}^{j}+\left( L^{k}\hat{\varrho}\left(
t\right) L_{k}^{\ast i}-\hat{\varrho}\left( t\right) K^{i}\right) \mathrm{d}%
\Lambda _{i}^{+},
\end{equation*}
with $K+K^{\ast }=L_{k}L^{k}$, $L^{k}=L_{k}^{\ast }$, $L_{k}^{\ast
i}=L_{i}^{k\ast }$ for any number of $k$'s, and arbitrary $K^{i}=K_{i}^{\ast
}$, $L_{j}^{k}$, $i,j=1,\ldots ,d$. This is the quantum stochastic
generalization of the general form \cite{Lin76} for the non-stochastic
(Lindblad) Master equation corresponding to the case $d=0$. In the case $d>0$
with pseudo-unitary block-matrix $\mathbf{L=}\left[ L_{\nu }^{\mu }\right]
_{\nu =-,\circ ,+}^{\mu =-,\circ ,+}$ in the sense $\mathbf{L}^{\star }=%
\mathbf{L}^{-1}$, it gives the general form of quantum stochastic Langevin
equation corresponding to the HP unitary evolution for $\chi \left( t\right) 
$ \cite{HuPa84}.

The nonlinear form of this decoherence equation for the exactly normalized
density operator $\hat{\rho}\left( t\right) =\hat{\varrho}\left( t\right) /%
\mathrm{Tr}_{\mathfrak{h}}\hat{\varrho}\left( t\right) $ was obtained for
different commutative It\^{o} algebras in \cite{Be90c, BaBe, Be92a}.

\medskip

\textbf{Acknowledgment:}

I would like to acknowledge the help of Robin Hudson and some of my students
attending the lecture course on Modern Quantum Theory who were the first who
read and commented on these notes containing the answers on some of their
questions. The best source on history and drama of quantum theory is in the
biographies of the great inventors, Schr\"{o}dinger, Bohr and Heisenberg 
\cite{Schr, Bohr, Heis}, and on the conceptual development of this theory
before the rise of quantum probability --\ in \cite{Jamm}. An excellent
essay ``The quantum age begins'', as well as short biographies with posters
and famous quotations of all mathematicians and physicists mentioned here
can be found on the mathematics website at St Andrews University --
http://www-history.mcs.st-and.ac.uk/history/, the use of which is
acknowledged.

\end{document}